\documentclass[pra]{revtex4}
\usepackage{eurosym}
\usepackage{amssymb}
\usepackage{graphicx}
\usepackage[dvipdfm]{hyperref}
\usepackage{verbatim}
\usepackage[normalem]{ulem}
\usepackage{color}
\usepackage{amsmath}

\begin{document}

\title{Suppression of Quantum-Mechanical Collapse in Bosonic Gases with
Intrinsic Repulsion: a Brief Review}
\author{B. A. Malomed}
\affiliation{Department of Physical Electronics, School of Electrical Engineering,
Faculty of Engineering, Tel Aviv University, Tel Aviv University, Tel Aviv
69978, Israel\\
malomed\@post.tau.ac.il}

\begin{abstract}
It is known that attractive potential $\sim -1/r^{2}$ gives rise to the
critical quantum collapse in the framework of the three-dimensional (3D)
linear Schr\"{o}dinger equation. This article summarizes theoretical
analysis, chiefly published in several original papers, which demonstrates
suppression of the collapse caused by this potential, and the creation of
the otherwise missing ground state in a 3D gas of bosonic dipoles pulled by
the same potential to the central charge, with repulsive contact
interactions between them, represented by the cubic term in the respective
Gross-Pitaevskii equation (GPE). In two dimensions (2D), quintic
self-repulsion is necessary for the suppression of the collapse;
alternatively, this may be provided by the effective quartic repulsion,
produced by the Lee-Huang-Yang correction to the GPE. 3D states carrying
angular momentum are constructed in the model with the symmetry reduced from
spherical to cylindrical by an external polarizing field. Interplay of the
collapse suppression and miscibility-immiscibility transition is considered
in a binary condensate. The consideration of the 3D setting in the form of
the many-body quantum system, with the help of the Monte Carlo method,
demonstrates that, although the quantum collapse cannot be fully suppressed,
the self-trapped states, predicted by the GPE, exist in the many-body
setting as metastable modes protected against the collapse by a tall
potential barrier.
\end{abstract}

\maketitle

quantum anomaly; ground state; self-trapping; Bose-Einstein condensate;
Gross-Pitaevskii equation; Thomas-Fermi approximation; mean-field
approximation; quantum phase transitions; Monte Carlo method

\section*{Abbreviations}

\noindent
\begin{tabular}{@{}ll}
2D & two-dimensional \\
3D & three-dimensional \\
BEC & Bose-Einstein condensate \\
GPE & Gross-Pitaevskii equation \\
LHY & Lee-Huang-Yang (correction to the mean-field theory) \\
GS & ground state \\
rms & root-mean-square (value) \\
TFA & Thomas-Fermi approximation%
\end{tabular}



\section{Introduction}

One of standard exercises given to students taking the course of quantum
mechanics is solving the three-dimensional (3D) Schr\"{o}dinger equation
with an isotropic attractive potential,
\begin{equation}
U(r)=-\frac{U_{0}}{2r^{2}},~U_{0}>0  \label{U}
\end{equation}%
\cite{LL}. This exercise offers a unique example of critical phenomena in
the nonrelativistic quantum theory. Indeed, the corresponding classical
(Newton's) equation of motion for the particle's coordinates, $\mathbf{r}%
=\left\{ x,y,z\right\} $,
\begin{equation}
\frac{d^{2}\mathbf{r}}{dt^{2}}=-\frac{\partial U}{\partial \mathbf{r}}\equiv
U_{0}\frac{\mathbf{r}}{r^{4}},  \label{class}
\end{equation}%
admits obvious rescaling, $t\equiv \tilde{t}/\sqrt{U_{0}}$, which eliminates
$U_{0}$ from Eq. (\ref{class}), thus making the solution invariant with
respect to the choice of a positive value of the potential strength, $U_{0}$%
. However, the invariance is lost in the corresponding 3D Schr\"{o}dinger
equation for wave function $\psi \left( \mathbf{r},t\right) $,%
\begin{equation}
i\psi _{t}=-\frac{1}{2}\nabla ^{2}\psi -\frac{U_{0}}{2r^{2}}\psi ,
\label{Schr}
\end{equation}%
in which $U_{0}$ cannot be removed by rescaling. This drastic difference
between the classical mechanical system and its quantum-mechanical
counterpart is known as the \textit{quantum anomaly} , alias
\textquotedblleft dimensional transmutation" \cite{anomaly,anomaly2}. The
consequence of the anomaly is well known: if an external trapping potential,
\begin{equation}
U_{\mathrm{trap}}=\frac{1}{2}\Omega ^{2}r^{2},  \label{trap}
\end{equation}%
is added to Eq. (\ref{Schr}), to make the integral norm,
\begin{equation}
N=\int \left\vert \psi (\mathbf{r})\right\vert ^{2}d\mathbf{r,}  \label{N}
\end{equation}%
convergent at $r\rightarrow \infty $, the Schr\"{o}dinger equation gives
rise to the normal set of trapped modes, starting from the ground state
(GS), at%
\begin{equation}
U_{0}<\left( U_{0}\right) _{\mathrm{cr}}^{(\mathrm{3D})}=1/4.  \label{1/4}
\end{equation}%
On the other hand, above this critical point, i.e., at $U_{0}>1/4$, the GS
does not exist (or, formally, speaking, it has an infinitely small size,
corresponding to energy $E\rightarrow -\infty $, which is known as
\textquotedblleft fall onto the center" \cite{LL}, the other name for which
is \textquotedblleft quantum collapse" \cite{anomaly,anomaly2}).

In the 2D space, the quantum collapse, driven by the same potential (\ref{U}%
), is more violent, taking place at any value $U_{0}>0$ (in other words, the
respective critical value is $\left( U_{0}\right) _{\mathrm{cr}}^{\mathrm{%
(2D)}}=0$). Finally, in the 1D case the same potential (\ref{U})\ gives rise
to a still stronger\textit{\ superselection} effect, which means splitting
the 1D space into two non-communicating subspaces, $x\gtrless 0$ \cite%
{superselection}.

A solution to the quantum-collapse problem in the 3D case was proposed in
terms of a linear quantum-field-theory, replacing the usual
quantum-mechanical wave function by the secondary-quantized field \cite%
{anomaly,anomaly2}. This approach makes it possible to introduce the GS,
which is missing at $U_{0}>1/4$ in the framework of standard quantum
mechanics. However, the solution does not predict a definite value of the
size of the newly created GS. Instead, the field-theory formulation, based
on the renormalization-group technique, introduces a GS with an \emph{%
arbitrary} spatial scale, in terms of which all other spatial sizes are
measured in that setting.

The present mini-review aims to summarize results produced by works which
elaborated another possibility to resolve the problem of the quantum
collapse. This possibility was proposed in Ref. \cite{HS1}, and then
developed, for more general situations, in works \cite{HS2} and \cite{HS3}.
The solution was based on the consideration of an ultracold gas of bosonic
particles pulled to the center by potential (\ref{U}). The gas was assumed
to be in the state of the Bose-Einstein condensate (BEC) \cite{BEC}, and the
suppression of the single-particle quantum collapse in this coherent
many-body setting was provided by repulsive contact interactions between
colliding particles in the gas. The solution was elaborated in the framework
of the mean-field approach \cite{BEC}, i.e., treating the single-particle
wave function, which represents all particles in the gas, as a classical
field governed by the corresponding Gross-Pitaevskii equation (GPE).

The same work \cite{HS1} offered a physical realization of potential (\ref{U}%
) in the 3D space, which was previously considered as a formal exercise \cite%
{LL}. The realization is provided by assuming that the bosonic particles are
small molecules carrying a permanent electric dipole moment, $d$, pulled by
the electrostatic force to a point-like charge, $Q$, placed at the origin,
which creates electric field $\mathbf{E}=Q\mathbf{r}/r^{3}$. In this
connection, it is relevant to mention that it has been demonstrated
experimentally that a free charge (ion), immersed in an ultracold gas, may
be kept at a fixed position by means of a laser-trapping technique \cite{ion}%
. Assuming that the orientation of the dipole carried by each particle is
locked to the local field, i.e., $\mathbf{d/}d=\mathrm{sgn}(Q)\left( \mathbf{%
r}/r\right) $, so as to minimize the interaction energy, the respective
interaction potential is $U(r)=-\mathbf{d}\cdot \mathbf{E}$, which is
tantamount to potential (\ref{U}) with strength
\begin{equation}
U_{0}=2|Q|d.  \label{U0}
\end{equation}%
As for the dipolar molecules which may be used to build the BEC under the
consideration, experimental results suggest that they may be, e.g., LiCs
\cite{LiCs} or KRb \cite{KRb}.

The gas of ultracold dipolar molecules, trapped in a \textit{pancake-shaped}
configuration shaped by an appropriate external potential \cite{2D-review},
with the central electric charge immersed in the gas as outlined above,
provides for the realization of the 2D version of the setting. An
alternative realization of the 2D setting is offered by a gas of polarizable
atoms without a permanent dielectric moment, while an effective moment is
induced in them by the electric field of a uniformly charged wire set
perpendicular to the pancake's plane \cite{Schmiedmayer}, or with an
effective magnetic moment induced by a current filament (e.g., an electron
beam) piercing the pancake perpendicularly.

In the context of 2D settings, it is relevant to mention that a quantum
anomaly was also predicted in a model described by the GPE in the 2D space,
for a gas of bosons with the repulsive contact interaction, trapped in the
harmonic-oscillator potential (\ref{trap}) \cite{Olshanii}. The anomaly
breaks the specific scaling invariance of this gas, which holds in the
mean-field approximation.

In terms of the GPE, the contact repulsive interaction in the bosonic gas is
represented by the cubic term \cite{BEC}. With the addition of this term,
and taking into regard the external trapping potential (\ref{trap}), which
is present in any experiment with ultracold atoms, the linear Schr\"{o}%
dinger equation (\ref{Schr}) is replaced by the GPE, which is written here
in the scaled form:
\begin{equation}
i\psi _{t}=-\frac{1}{2}\left( \nabla ^{2}+\frac{U_{0}}{r^{2}}-\Omega
^{2}r^{2}\right) \psi +\left\vert \psi \right\vert ^{2}\psi .  \label{GPE}
\end{equation}

It is relevant to mention that the 3D GPE with the self-attractive
interaction, which corresponds to the opposite sign in front of the cubic
term in Eq. (\ref{GPE}), gives rise, in the absence of the attractive
potential ($U_{0}=0$) to the well-known supercritical wave collapse \cite%
{Berge'}. A relation of this setting to Eq. (\ref{GPE}) is that the
inclusion of the trapping potential $\sim \Omega ^{2}$ gives rise to \emph{%
stable} bound states in the form of spherically symmetric bound states and
ones with vorticity $m=1$ (cf. Eq. (\ref{psi}) below), provided that norm $N$
does not exceed a certain critical value \cite{Dodd}-\cite{DumDum}.

The energy (Hamiltonian) corresponding to Eq. (\ref{GPE}) is%
\begin{equation}
E=\frac{1}{2}\int \left[ \left\vert \nabla \psi \right\vert ^{2}-\left(
\frac{U_{0}}{r^{2}}-\Omega ^{2}r^{2}\right) \left\vert \psi \right\vert
^{2}+\left\vert \psi ^{4}\right\vert \right] d\mathbf{r},  \label{E}
\end{equation}

The scaled variables and constants, in terms of which Eq. (\ref{GPE}) is
written, are related to their counterparts measured in physical units:%
\begin{equation}
\mathbf{r}=\frac{\mathbf{r}_{\mathrm{ph}}}{r_{0}},~t=\frac{\hbar }{mr_{0}^{2}%
}t_{\mathrm{ph}},~~\psi =2\sqrt{\pi a_{s}}r_{0}\psi _{\mathrm{ph}},~U_{0}=%
\frac{m}{\hbar ^{2}}\left( U_{0}\right) _{\mathrm{ph}},~\Omega =\frac{%
mr_{0}^{2}}{\hbar }\Omega _{\mathrm{ph}},  \label{ph}
\end{equation}%
where $m$ and $a_{s}$ are the bosonic mass and \textit{s}-scattering length,
which accounts for the repulsive interactions between the particles \cite%
{BEC}, and $r_{0}$ is an arbitrary spatial scale. The total number of bosons
in the gas is given by%
\begin{equation}
N_{\mathrm{ph}}=\int \left\vert \psi _{\mathrm{ph}}(\mathbf{r}_{\mathrm{ph}%
})\right\vert ^{2}d\mathbf{r}_{\mathrm{ph}}\equiv \frac{r_{0}N}{4\pi a_{s}},
\label{Nph}
\end{equation}%
where the norm of the scaled wave function is given by Eq. (\ref{N}).

It is relevant to note that, as it follows from Eq. (\ref{U0}) and rescaling
(\ref{ph}), the above-mentioned critical value, $U_{0}=1/4$, of the strength
of the attractive potential (see Eq. (\ref{1/4})) corresponds to a very
small dipole moment, $d\sim 10^{-6}$ Debye, if central charge $Q$ is taken
as the elementary charge, and the mass of the particle is $\sim 100$ proton
masses. Therefore, the overcritical case of $U_{0}>1/4$, the consideration
of which is the main objective of the present article, is relevant in the
actual physical context.

Taken as Eq. (\ref{GPE}), the GPE neglects dipole-dipole interactions
between the particles. These interactions can be taken into account in the
framework of another application of the mean-field approach. Indeed, the
local density of the dipole moment in the gas (i.e., the dielectric
polarization of the medium) is $\mathbf{P}=\mathbf{d}\left\vert \psi (%
\mathbf{r})\right\vert ^{2}$, hence the electrostatic field generated by the
polarization, $\mathbf{E}_{d}$, is determined by the Poisson equation, $%
\nabla \cdot \left( \mathbf{E}_{d}+4\pi \mathbf{P}\right) =0$, which can be
solved immediately:%
\begin{equation}
\mathbf{E}_{d}=-4\pi \mathbf{P}\equiv -4\pi \mathbf{d}\left\vert \psi (%
\mathbf{r})\right\vert ^{2}.  \label{EP}
\end{equation}%
Then, the extra term in the GPE, accounting for the interaction of the local
dipole with the collective field (\ref{EP}), created by all the other
dipoles, is%
\begin{equation}
-\left( \mathbf{d}\cdot \mathbf{E}_{d}\right) \psi \equiv 4\pi
d^{2}\left\vert \psi \right\vert ^{2}\psi .  \label{dip}
\end{equation}%
This term, if added to Eq. (\ref{GPE}), may be absorbed into a redefinition
of the scattering length accounting for the repulsion between the particles.
In the underlying physical units, this amounts to
\begin{equation}
a_{s}\rightarrow \left( a_{s}\right) _{\mathrm{eff}}\equiv
a_{s}+md^{2}/\hbar ^{2},  \label{eff}
\end{equation}%
where $m$ is the mass of the dipolar molecule. For the typical value of $%
a_{s}\sim 10$ nm and the above-mentioned mass of the particle, $\sim 100$
proton masses, Eq. (\ref{eff}) demonstrates that the additional term is
essential for dipole moments $d\gtrsim 0.3$ Debye.

The rest of the article is organized as follows. In Section II, results are
reported for the basic model, outlined above, as per Ref. \cite{HS1}.
Particular subsections of Section II first recapitulate the description of
the 3D and 2D collapse in the framework of Schr\"{o}dinger equation (\ref%
{Schr}), which includes the trapping potential (\ref{trap}), and then
present main results obtained in the 3D case on the basis of Eq. (\ref{GPE})
(with $\Omega =0$, as the trapping potential is not a necessary ingredient
of the nonlinear model, on the contrary to the linear one). The results
explicitly demonstrate the creation of the originally missing GS by the
self-repulsive cubic nonlinearity at $U_{0}>1/4$. In addition, a subsection
of Section II reports a new result, \textit{viz}., a quantum phase
transition in the GS of the model which includes the Lee-Huang-Yang (LHY)
correction \cite{LHY} to the mean-field GPE.\ The summary of results for the
2D nonlinear model are also presented in Section II. It is demonstrated that
the cubic self-repulsive term is insufficient for the suppression of the 2D
quantum collapse and restoration of the missing GS. This is possible if a
\emph{quintic} repulsive term is included in the GPE, which may account for
three- body collisions, or if the quartic LHY correction is included in the
effective two-dimensional GPE . A short subsection concluding Section II
formulates a challenging problem of the consideration of the quantum
collapse in the gas of fermions.

Section III addresses, along the lines of Ref. \cite{HS2}, the collapse
suppression and creation of the GS in the 3D model with the symmetry of the
effective attractive potential reduced from spherical to cylindrical by an
external field which polarizes dipole moments of the particles. In this
version of the model, states carrying the angular momentum are constructed
too, in addition to the GS.

Section IV deals with a two-component model in 3D, which makes it possible
to consider the interplay of the collapse suppression and the transition
between miscibility and immiscibility in the binary system. A weak quantum
phase transition, which occurs in that setting, is also briefly considered
in Section IV.

Section V presents results for the basic 3D model, considered in terms of
the many-body\ quantum theory, as per Ref. \cite{GEA}, with the help of
variational approximation for the many-body wave functions and numerically
implemented Monte Carlo method. The main result is that, strictly speaking,
the quantum collapse is not fully suppressed in the many-body theory, but,
nevertheless, the noncollapsing self-trapped state, predicted by the
mean-field theory, exists as a metastable one, insulated from the collapse
by a tall potential barrier.

The paper is concluded by Section VI, which also suggests directions for
further work on this general topic.

\section{The basic three- and two-dimensional models}

This section summarizes results produced in Ref. \cite{HS1}. The quantum
phase transition driven by the LHY correction to the mean-field theory,
briefly outlined in subsection 3, is a new finding.

\subsection{The quantum collapse in the linear Schr\"{o}dinger equation}

First, it is relevant to recapitulate the analysis of linear Schr\"{o}dinger
equation (\ref{Schr}), to which the trapping potential (\ref{trap}) is
added:
\begin{equation}
i\psi _{t}=-\frac{1}{2}\left( \nabla ^{2}+\frac{U_{0}}{r^{2}}-\Omega
^{2}r^{2}\right) \psi .  \label{linear}
\end{equation}%
Stationary solutions of Eq. (\ref{linear}) in 3D spherical coordinates, $%
\left( r,\theta ,\varphi \right) $, are looked for as
\begin{equation}
\psi _{\mathrm{3D}}=\exp (-i\mu t)Y_{lm}\left( \theta ,\varphi \right) \Phi
(r),  \label{psi}
\end{equation}%
where $\mu $ is the energy eigenvalue (or chemical potential, in terms of
the GPE), $Y_{lm}\left( \theta ,\varphi \right) $ is the spherical harmonic
with quantum numbers $\left( l,m\right) $, and radial wave function $\Phi
(r) $ is real. Substituting ansatz (\ref{psi}) in Eq. (\ref{linear}), two
\emph{exact} solutions for $\Phi (r)$ can be found:
\begin{gather}
\Phi (r)=\Phi _{0}r^{-\sigma _{\pm }}\exp \left( -\Omega r^{2}/2\right) ,
\label{exact} \\
\mu =\Omega \left( \frac{3}{2}-\sigma _{\pm }\right) ,~\sigma _{\pm }\equiv
\frac{1}{2}\pm \sqrt{\frac{1}{4}-U_{l}},  \label{sigma3D}
\end{gather}%
which exist under condition%
\begin{equation}
U_{l}\equiv U_{0}-l\left( l+1\right) <1/4.  \label{Ul}
\end{equation}%
The smaller value of $\mu $ (in the case of $l=0$, it defines the GS of the
system under the consideration) corresponds to $\sigma _{+}$, i.e.,, the top
sign in Eq. (\ref{sigma3D}). The wave function is characterized by its norm (%
\ref{N}),
\begin{equation}
N=4\pi \int_{0}^{\infty }\Phi ^{2}(r)r^{2}dr=2\pi \Phi _{0}^{2}\Omega
^{-\left( 1\mp \sqrt{\frac{1}{4}-U_{l}}\right) }\Gamma \left( 1\mp \sqrt{%
\frac{1}{4}-U_{l}}\right) ,  \label{N3D}
\end{equation}%
where $\Gamma $ is the Gamma-function. Equation (\ref{N3D}) shows why the
trapping potential $\sim \Omega ^{2}$ is necessary for the existence of
physically relevant (normalizable) eigenmodes of the linear Schr\"{o}dinger
equation (\ref{linear}), as norm (\ref{N3D}) diverges (due to its weak
localization at $r\rightarrow \infty $) in the limit of $\Omega \rightarrow
0 $.

\begin{figure}[t]
\begin{center}
\includegraphics[height=4.cm]{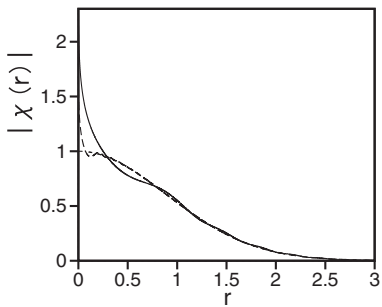}
\end{center}
\caption{Radial profiles of $|\protect\chi (r,t)|\equiv \protect\sqrt{r}|%
\protect\psi (r)|$ at $t=0$, $0.005$ and $0.1$ (dotted, dashed, and solid
curves, respectively), as originally produced in Ref. \protect\cite{HS1} by
simulations of Eq. (\protect\ref{linear}) with $\Omega ^{2}=0.1$ and $%
U_{0}=0.27$, which slightly exceeds the critical one, $\left( U_{0}\right) _{%
\mathrm{cr}}^{\mathrm{(3D)}}=1/4$. The initial conditions is taken as $%
\protect\psi _{0}(r)=r^{-1/2}\exp (-\Omega r^{2}/2)$, which is the exact
stationary wave function for $U_{0}=1/4$, i.e., precisely at the critical
point, taken as per Eqs. (\protect\ref{exact}) and (\protect\ref{sigma3D})
(for this reason, the evolution of the wave function is displayed here in
terms of $\protect\sqrt{r}|\protect\psi (r)|$). The simulations demonstrate
the onset of the quantum collapse in the linear Schr\"{o}dinger equation.}
\label{fig1}
\end{figure}

These solutions for the stationary wave functions do not exist at $U_{l}>1/4$
(note that the presence of the angular momentum, $l\geq 1$, secures the
existence of the bound states at essentially larger values of $U_{0}$, as
per Eq. (\ref{Ul})). The nonexistence of stationary wave functions implies
that the system suffers the onset of the quantum collapse, as confirmed by
simulations of time-dependent equation (\ref{linear}), see an example in
Fig. \ref{fig1}. Indeed, a set of instantaneous profiles of $\sqrt{r}|\psi
(r,t)|$, shown in Fig. \ref{fig1} for the weakly overcritical case, $%
U_{0}=0.27$, with $l=0$, confirm the development of the self-compression
(finally, collapse) of the wave function towards $r=0$ (in the simulations,
the collapse is eventually arrested due to a finite mesh size of the
numerical scheme).

In 2D, the GS solution to Eq. (\ref{linear}) exists only for $U_{0}<0$. In
the exact form, the GS wave function is given by Eqs. (\ref{psi}) and (\ref%
{exact}), but with (\ref{sigma3D}) replaced by
\begin{equation}
\mu =\Omega \left( 1-\sigma _{\pm }\right) ,\sigma _{\pm }=\pm \sqrt{-U_{0}}.
\label{sigma2D}
\end{equation}%
Direct simulations of the 2D equation (\ref{linear}) at $U_{0}>0$ also
demonstrate the onset of the collapse dynamics.

\subsection{The three-dimensional ground state (GS) created by the cubic
self-repulsive nonlinearity}

The most essential results may be produced by GPE (\ref{GPE}) without an
external trapping potential, hence the equation simplifies to%
\begin{equation}
i\psi _{t}=-\frac{1}{2}\left( \nabla ^{2}+\frac{U_{0}}{r^{2}}\right) \psi
+\left\vert \psi \right\vert ^{2}\psi .  \label{GPE2}
\end{equation}%
The substitution of $\psi =e^{-i\mu t}\Phi (r)$ for isotropic stationary
states of Eq. (\ref{GPE2}) (here, solely $l=0$ is considered, cf. Eq. (\ref%
{psi}), with the intention to construct the GS, which always has $l=0$)
yields equation
\begin{equation}
\mu \Phi =-\frac{1}{2}\left( \frac{d^{2}\Phi }{dr^{2}}+\frac{2}{r}\frac{%
d\Phi }{dr}+\frac{U_{0}}{r^{2}}\right) \Phi +\Phi ^{3}.  \label{phi1}
\end{equation}%
Scaling invariance of Eq. (\ref{phi1}) at $r\rightarrow 0$ suggests that the
respective asymptotic form of the solution should be $\Phi \sim 1/r$,
therefore  solutions are looked for as
\begin{equation}
\Phi \left( r\right) =\frac{\chi (r)}{r},  \label{psichi3D}
\end{equation}%
with function $\chi (r)$ obeying equation
\begin{equation}
\mu \chi =-\frac{1}{2}\left[ \chi ^{\prime \prime }+\left( \frac{U_{0}}{r^{2}%
}-\Omega ^{2}r^{2}\right) \chi \right] +\frac{\chi ^{3}}{r^{2}}.  \label{chi}
\end{equation}

Asymptotic forms of solutions to Eq. (\ref{chi}) can be readily constructed
for $r\rightarrow 0$ and $r\rightarrow \infty $. First, the expansion at $%
r\rightarrow 0$ yields
\begin{equation}
\chi (r)=\sqrt{U_{0}/2}+\chi _{1}r^{s/2},~s\equiv 1+\sqrt{1+8U_{0}},
\label{r=0}
\end{equation}%
where $\chi _{1}$ is a free constant, in terms of this expansion. At $%
r\rightarrow \infty $ the asymptotic form of the bound-state solution with $%
\mu <0$ is
\begin{equation}
\chi =\chi _{0}\exp \left( -\sqrt{-2\mu }r\right) ,  \label{r=infty}
\end{equation}%
where $\chi _{0}$ is an arbitrary constant, in terms of the expansion for $%
r\rightarrow \infty $. A global analytical approximation can be constructed
as an interpolation, stitching together the asymptotic forms (\ref{r=0})
(where the correction term $\sim \chi _{1}$ is neglected, in the present
approximation) and (\ref{r=infty}):{\
\begin{equation}
\psi (r,t)=\sqrt{\frac{U_{0}}{2}}e^{-i\mu t}r^{-1}e^{-\sqrt{-2\mu }r}.~
\label{inter}
\end{equation}%
}

{Note that the singularity of wave function (\ref{inter}) at }$r\rightarrow 0
$ is acceptable, as the respective integral (\ref{N}) converges at small $r$%
. It is also relevant to mention that, following the substitution of the
asymptotic form (\ref{inter}) in the effective \textit{pseudopotential} in
Eq. (\ref{GPE2}), which includes the nonlinear term, $U_{\mathrm{pseudo}%
}(r)\equiv -(1/2)U_{0}r^{-2}+\left\vert \psi (r)\right\vert ^{2}$, the
singularity $\sim r^{-2}$ at $r\rightarrow 0$ cancels out in it. Note also
that a more singular attractive potential, $U(r)=-U_{0}/r^{b}$, with $U_{0}>0
$ and $b>2$, gives rise to asymptotic form $|\psi |^{2}\approx U_{0}/r^{b}$
of the solution at $r\rightarrow 0$, hence the corresponding norm still
converges at $b<3$.

{Due to the nonlinearity of Eq. (\ref{GPE2}), the chemical potential of the
GS depends on its norm. Using approximation (\ref{inter}), it is easy to
calculate }$\mu $ as a function of $N$:%
\begin{equation}
\mu =-\frac{1}{2}\left( \frac{\pi U_{0}}{N}\right) ^{2}.  \label{mu}
\end{equation}%
In fact, scaling $\mu \sim N^{-2}$ is an exact property\emph{\ }of solutions
to Eq. (\ref{GPE2}), which follows from a straightforward analysis of this
equation. Note also that, in the limit of $\mu \rightarrow -0$, Eq. (\ref%
{inter}) gives a particular \emph{exact} solution of Eq. (\ref{GPE2}),
\begin{equation}
\psi _{\mu =0}(r)=\sqrt{U_{0}/2}r^{-1},  \label{mu=0}
\end{equation}%
although its norm diverges.

\begin{figure}[t]
\begin{center}
\includegraphics[height=4.cm]{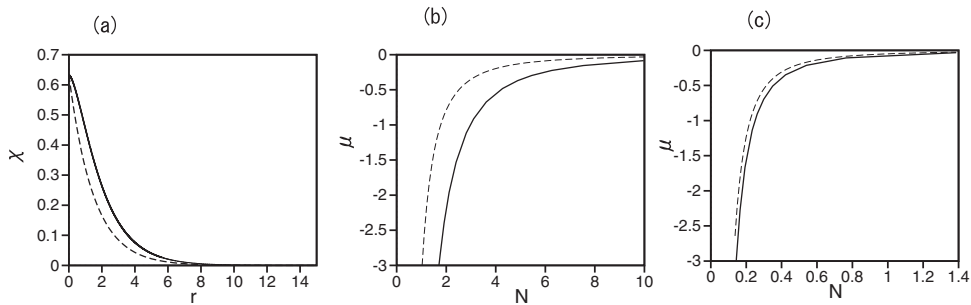}
\end{center}
\caption{ (a) A typical example of the 3D ground state, shown in terms of $%
\protect\chi (r)\equiv r\left\vert \protect\psi (r)\right\vert $, produced
by the GPE (\protect\ref{GPE2}), as per Ref. \protect\cite{HS1}, without the
external trap ($\Omega =0$), for $U_{0}=0.8$ and $\protect\mu =-0.225$.
Panels (b) and (c) display curves $\protect\mu (N)$ for the ground-state
families with $U_{0}=0.8$ and $0.1$. These strengths of the attractive
potential are, respectively, larger and smaller than the critical value $1/4$
(see Eq. (\protect\ref{1/4})) for the linear Schr\"{o}dinger equation (%
\protect\ref{linear}). Here, solid and dashed curves depict, respectively,
the numerical results and analytical approximation given by Eqs. (\protect
\ref{inter}) and (\protect\ref{mu}) (in panels (b) and (c), the curves
follow scaling $\protect\mu \sim N^{-2}$, which is an exact property of Eq. (%
\protect\ref{GPE2})). In particular, the analytical approximation predicts $%
N(\protect\mu =-0.225)=5.30$ for $U_{0}=0.8$ (the solution shown in (a)),
while the numerically found counterpart of this value is $N_{\mathrm{num}}(%
\protect\mu =-0.225)=6.26$. The convergence of the numerical and analytical
curves for $N(\protect\mu )$ at $\protect\mu \rightarrow 0$ corresponds to
the fact that Eq. (\protect\ref{inter}) gives exact solution (\protect\ref%
{exact})\ in this limit.}
\label{fig2}
\end{figure}

Equation (\ref{chi}) can be easily solved in a numerical form. A typical
example of the numerical GS solution, along with approximation (\ref{inter}%
), is displayed in Fig. \ref{fig2}(a) for $U_{0}=0.8$, which is essentially
\emph{larger} than the critical value of the attraction strength, $\left(
U_{0}\right) _{\mathrm{cr}}^{(\mathrm{3D})}=1/4$ (see Eq. (\ref{1/4})),
beyond which linear Schr\"{o}dinger equation (\ref{linear}) has no GS.
Further, Figs. \ref{fig2}(b) and \ref{fig2}(c) represent the family of the
GS\ states, by means of dependences $\mu (N)$, for two values, $U_{0}=0.8$
and $0.1$, which are, respectively, larger and smaller than $1/4$. Thus, on
the contrary to the linear Schr\"{o}dinger equation, GPE (\ref{GPE2})
maintains the GS at all values of $U_{0}$ and $N$. In other words, the
inclusion of the repulsive cubic term in Eq. (\ref{GPE2}) completely
suppresses the quantum collapse in the 3D space, creating the GS where it
does not exist in the linear Schr\"{o}dinger equation.

The analytical approximation (\ref{inter}) suggests an estimate for the
radial size of the GS created by the repulsive nonlinearity:%
\begin{equation}
r_{\mathrm{GS}}^{2}\equiv \frac{4\pi }{N}\int_{0}^{\infty }\left\vert \psi
(r)\right\vert ^{2}r^{4}dr=\frac{N^{2}}{2\pi ^{2}U_{0}^{2}}.  \label{R}
\end{equation}%
It is relevant to rewrite this estimate in terms of physical units, as per
Eqs. (\ref{ph}), (\ref{Nph}), and (\ref{eff}):%
\begin{equation}
\left( r_{\mathrm{GS}}\right) _{\mathrm{ph}}\equiv r_{0}r_{\mathrm{GS}}=%
\frac{2\sqrt{2}\left( \hbar ^{2}a_{s}+md^{2}\right) N_{\mathrm{ph}}}{m\left(
U_{0}\right) _{\mathrm{ph}}}.  \label{Rph}
\end{equation}%
Note that arbitrary spatial scale $r_{0}$, which was used in rescalings (\ref%
{ph}) and (\ref{Nph}), cancels out in Eq. (\ref{Nph}). Thus, GPE (\ref{GPE2}%
) uniquely predicts the radius of the restored GS, in terms of physical
parameters of the model.

It is natural that $r_{\mathrm{GS}}$, given by Eq. (\ref{Rph}), shrinks to
zero in the limit of vanishing nonlinearity, which is tantamount to $N_{%
\mathrm{ph}}\rightarrow 0$, implying the onset of the collapse in the
framework of the linear Schr\"{o}dinger equation. Note also too that, if the
contribution from by the dipole-dipole interactions ($\sim d^{2}$) dominates
over the contact interactions in Eq. (\ref{Rph}) ($md^{2}\gtrsim \hbar
^{2}a_{s}$), the latter result strongly simplifies, taking into regard Eq. (%
\ref{U0}): $\left( r_{\mathrm{GS}}\right) _{\mathrm{ph}}=\left( \sqrt{2}%
d/|Q|\right) N_{\mathrm{ph}}$. Then, for $Q$ equal to the elementary charge,
$d\sim 1$ Debye, and $N_{\mathrm{ph}}\sim 10^{5}$, the latter estimate
predicts the GS with radius $\sim 3$ $\mathrm{\mu }$m. This result upholds
the self-consistency of the model, as the mean-filed approximation (and the
respective GPE) are definitely applicable for scales $\gtrsim 1$ $\mathrm{%
\mu }$m \cite{BEC}.

It is worthy to stress that Eq. (\ref{GPE}), which does not include the
trapping potential, predicts the GS with the \emph{finite norm} at $%
U_{0}<1/4 $, as the norm of the corresponding stationary solutions to the
linear equation (\ref{linear}), see Eqs. (\ref{psi}) and (\ref{exact}),
diverges at $\Omega =0$. Lastly, simulations of Eq. (\ref{GPE}) with random
perturbations added to the stationary solutions demonstrate that the GS is
always dynamically stable \cite{HS1}. The stability agrees too with the
\textit{anti-Vakhitov-Kolokolov} criterion, $d\mu /dN>0$, which is a
necessary stability condition for localized states supported by
self-repulsive nonlinearities \cite{anti} (the original Vakhitov-Kolokolov
criterion, $d\mu /dN<0$, is the necessary stability condition in the case of
self-attraction \cite{VK,Berge'}) .

\subsection{The quantum phase transition induced by the Lee-Huang-Yang
(LHY)\ correction to the mean-field theory}

The singularity $\sim r^{-1}$ of the stationary wave function at $%
r\rightarrow 0$, as seen in Eqs. (\ref{psichi3D}) and (\ref{r=0}), suggests
that, although the singularity is integrable, as the respective 3D integral
for the total norm converges, the LHY correction \cite{LHY} to the
mean-field theory, which is relevant for higher values of the condensate's
density, must be taken into regard. As shown in Ref. \cite{Petrov,AP}, the
scaled GPE with this correction, represented by coefficient $g_{\mathrm{LHY}%
}>0$, is%
\begin{equation}
i\psi _{t}=-\frac{1}{2}\left( \nabla ^{2}+\frac{U_{0}}{r^{2}}\right) \psi
+\left\vert \psi \right\vert ^{2}\psi +g_{\mathrm{LHY}}\left\vert \psi
\right\vert ^{3}\psi .  \label{+LHY}
\end{equation}%
Then, the asymptotic form of the stationary wave function at $r\rightarrow 0$%
, which was found above in the form determined by the cubic term in Eq. (\ref%
{GPE2}), $\Phi (r)\approx \sqrt{U_{0}/2}r^{-1}$, is replaced by%
\begin{equation}
\Phi _{\mathrm{LHY}}(r)\approx \left( U_{0}/2-1/9\right) ^{1/3}r^{-2/3},
\label{-2/3}
\end{equation}%
under the condition of $U_{0}>2/9$, while at $0<U_{0}<2/9$ the asymptotic
form is determined by the linearization of Eq. (\ref{+LHY}), leading to the
same result as given by Eqs. (\ref{exact}) and (\ref{sigma3D}) with $\sigma
=\sigma _{-}$ (and $U_{l}$ replaced by $U_{0}$):%
\begin{equation}
\Phi _{\mathrm{LHY}}(r)\approx \Phi _{0}r^{-\left( 1/2-\sqrt{1/4-U_{0}}%
\right) },  \label{sigma-}
\end{equation}%
where $\Phi _{0}$ is an arbitrary constant in terms of the expansion at $%
r\rightarrow 0$. Note that the wave function with asymptotic form $%
r^{-\sigma _{+}}$, which corresponds to the GS in the linear Schr\"{o}dinger
equation, is incompatible with the presence of the LHY term in Eq. (\ref%
{+LHY}), although power $-2/3$ in expression (\ref{+LHY}) coincides, at the
critical point, $U_{0}=2/9$, with $\sigma _{+}$, rather than $\sigma _{-}$.

Thus, the jump from the asymptotic form (\ref{sigma-}), produced by the
linear Schr\"{o}dinger equation, to one (\ref{-2/3}) generated by the LHY
term at $U_{0}=2/9$ (in particular, the jump between $\sigma _{-}$ and $%
\sigma _{+}$), takes place at $U_{0}=2/9$, which is a signature of a \textit{%
quantum phase transition}. Examples of such phase transitions were studied
in many-body settings \cite{Grisha} and in many other systems \cite{other1}-%
\cite{other5}.

Lastly, the LHY term may stabilize the bosonic gas pulled to the center by
potential (\ref{U}) even in the case of the effective attractive
interaction, which is possible in a binary condensate, with intrinsic
self-repulsion in each components, and dominating attraction between them,
as proposed in Refs. \cite{Petrov} and \cite{AP}, and realized
experimentally in the form of \textquotedblleft quantum droplets" (in the
binary condensate of $^{39}$K) in Refs. \cite{droplet1}-\cite{droplet3}. For
the symmetric mixture, with equal wave functions of the two components, the
effective GPE takes the form of Eq. (\ref{+LHY}) with the opposite sign in
front of the cubic term. This model may be a subject for special
consideration.

\subsection{The two-dimensional ground state created by the quintic
self-repulsive nonlinearity}

As mentioned above, the GPE in the form of Eq. (\ref{GPE}) may be relevant,
as a physical model, in 2D too. However, the 2D version of norm (\ref{N}) of
the wave function with asymptotic form $\sim r^{-1}$ at $r\rightarrow 0$,
which follows from this equation (see Eq. (\ref{inter})), logarithmically
diverges at small $r$. This means that, the cubic self-repulsion is not
strong enough to suppress the collapse in the 2D geometry. On the other
hand, the GPE may also include the quintic repulsive term accounting for
three-body collisions, provided that the collisions do not give rise to
conspicuous losses \cite{3-body1,3-body2}.

The 2D GPE can be derived from the underlying 3D version if tight
confinement, with the respective harmonic-oscillator length, $a_{\perp }$,
is imposed in direction $z$ by the trapping harmonic-oscillator potential,
reducing the effective dimension to that of the plane with remaining
coordinates $\left( x,y\right) $ \cite{Shlyap}-\cite{Delgado}. In
particular, if the dominating quintic terms appears in the 3D GPE with
coefficient $g_{5}$, the reduction to the 2D equation replaces it by $\left(
\sqrt{3}\pi a_{\perp }^{2}\right) ^{-1}g_{5}$.

In the scaled form, the 2D equation it written in the polar coordinates, $%
\left( r,\theta \right) $, as
\begin{equation}
i\psi _{t}=-\frac{1}{2}\left( \psi _{rr}+\frac{1}{r}\psi _{r}+r+\frac{1}{%
r^{2}}\psi _{\theta \theta }+\frac{U_{0}}{r^{2}}\right) \psi +\left\vert
\psi \right\vert ^{4}\psi .  \label{psi2d}
\end{equation}%
Stationary solutions to Eq. (\ref{psi2d}) (not only the GS, but also for
states carrying the angular momentum) are looked for as%
\begin{equation}
\psi _{\mathrm{2D}}\left( r,t\right) =e^{-i\mu t+il\theta }r^{-1/2}\chi (r),
\label{psichi2D}
\end{equation}%
where integer $l$ is the azimuthal quantum number, cf. Eq. (\ref{psichi3D}).
The substitution of this ansatz in Eq. (\ref{psi2d}) yields an equation for
real $\chi _{\mathrm{2D}}(r)$:%
\begin{equation}
\mu \chi =-\frac{1}{2}\left[ \chi ^{\prime \prime }+\left( U_{l}^{\mathrm{%
(2D)}}+\frac{1}{4}\right) r^{-2}\chi \right] +r^{-2}\chi ^{5},  \label{chi2D}
\end{equation}%
with $U_{l}^{\mathrm{(2D)}}\equiv U_{0}-l^{2}$, cf. Eq. (\ref{Ul}). Note
that, unlike the 3D case, in 2D nonlinear model the analysis is possible
equally well for $l=0$ (the GS) and $l\geq 1$.

The expansion of the solution to Eq. (\ref{chi2D}) at $r\rightarrow 0$
yields
\begin{equation}
\chi =\left[ \frac{1}{2}\left( U_{l}^{\mathrm{(2D)}}+\frac{1}{4}\right) %
\right] ^{1/4}+\chi _{1}r^{s},  \label{r=0-2D}
\end{equation}%
where $s=\left( 1/2\right) \left( 1+\sqrt{5+16U_{l}^{\mathrm{(2D)}}}\right) $%
, and $\chi _{1}$ is an arbitrary constant in terms of this expansion, cf.
Eq. (\ref{r=0}) in the 3D case. The solution with a finite norm exists at $%
U_{l}^{\mathrm{(2D)}}>-1/4$, representing, at $U_{l}^{\mathrm{(2D)}}>0$, the
suppression of the collapse and creation of the GS ($l=0$), or the state
with $l\geq 1$, by the quintic self-repulsive term.

Combining the 2D asymptotic form (\ref{r=0-2D}), valid at $r\rightarrow 0$,
and the obvious approximation valid at $r\rightarrow \infty $, $\chi _{%
\mathrm{2D}}\approx \chi _{0}\exp \left( -\sqrt{-2\mu }r\right) $, one
derives an interpolation formula for the GS, and the dependence $\mu (N)$
following from it, cf. Eqs. (\ref{inter}) and (\ref{mu}) in the 3D case:
\begin{eqnarray}
\psi _{\mathrm{2D}}\left( r,t\right) &=&\left[ \frac{1}{2}\left( U_{l}^{%
\mathrm{(2D)}}+\frac{1}{4}\right) \right] ^{1/4}e^{-i\mu t+il\theta
}r^{-1/2}e^{-\sqrt{-2\mu }r},  \notag \\
\mu &=&-\left( U_{l}^{\mathrm{(2D)}}+\frac{1}{4}\right) \left( \frac{\pi }{2N%
}\right) ^{2},  \label{inter2D}
\end{eqnarray}%
Similar to the situation in the 3D case, Eq. (\ref{inter2D}) gives an exact
wave function with a divergent norm in the limit of $\mu \rightarrow -0$,%
\begin{equation}
\psi _{\mathrm{2D}}^{(\mu =0)}(r)=\left[ \frac{1}{2}\left( U_{l}^{\mathrm{%
(2D)}}+\frac{1}{4}\right) \right] ^{1/4}e^{il\theta }r^{-1/2},
\label{mu=0_2D}
\end{equation}%
cf. Eq. (\ref{mu=0}).

The approximation (\ref{inter2D}) makes it possible to define the rms radial
size of the two-dimensional GS, cf. Eq. (\ref{R}) in the 3D case:
\begin{equation}
r_{\mathrm{GS}}^{\mathrm{(2D)}}\equiv \sqrt{\frac{2\pi }{N}\int_{0}^{\infty
}\left\vert \psi _{\mathrm{2D}}(r)\right\vert ^{2}r^{3}dr}=\frac{N}{\pi
\sqrt{\left( U_{l}^{\mathrm{(2D)}}+1/4\right) }}.  \label{R2D}
\end{equation}

Note that the quintic term supports the GS in 2D even at $0<-U_{l}^{\mathrm{%
(2D)}}<1/4$, when the central potential is \emph{repulsive}. The correctness
of this counter-intuitive conclusion is corroborated by the above-mentioned
fact that the analytical approximation (\ref{inter2D}) gives exact solution (%
\ref{mu=0_2D}) for $\mu \rightarrow 0$, including the case of $0<-U_{l}^{%
\mathrm{(2D)}}<1/4$.

\begin{figure}[t]
\begin{center}
\includegraphics[height=4cm]{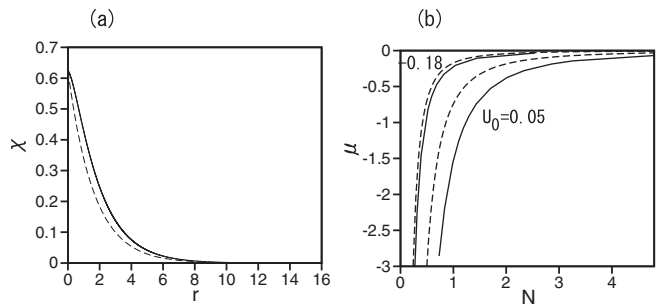}
\end{center}
\caption{(a) The radial profile of the ground state in the 2D model with the
quintic nonlinearity, for $U_{0}=0.05$ and $\protect\mu =-0.1867$. (b)
Curves $\protect\mu (N)$ for the ground states with $U_{0}=-0.18$ and $%
U_{0}=0.05$. In both panels (shown as per Ref. \protect\cite{HS1}), the
numerical results and the respective analytical approximation (\protect\ref%
{inter2D}) are depicted by the continuous and dashed curves, respectively.
The convergence of the numerical and analytical curves for $N(\protect\mu )$
at $\protect\mu \rightarrow -0$ corresponds to the fact that Eq. (\protect
\ref{inter2D}) gives the exact solution (\protect\ref{mu=0_2D}) in this
limit.}
\label{fig3}
\end{figure}

An example of the stable GS, and curves $\mu (N)$ for the GS in 2D are
displayed, along with the analytical approximation (\ref{inter2D}), in Fig. %
\ref{fig3} (referring to $l=0$, although it actually makes no difference in
the plots). The $\mu (N)$ curves are shown for both signs of the central
potential, $U_{0}=-0.18$ and $U_{0}=0.05$. Simulations of the perturbed
evolution in the framework of Eq. (\ref{psi2d}) confirm stability of the GS
families.

Generally, the results for the 2D model are more formal than those
summarized above for 3D, as the realization of the dominant quintic
nonlinearity in BEC is problematic, in experimentally relevant settings. On
the other hand, the LHY correction to the GPE is also sufficient to provide
the suppression of the quantum collapse and restoration of the GS in the 2D
setting. The same dimension-reduction procedure as outlined above, will
replace the original LHY coefficient in 3D equation (\ref{+LHY}) by $\sqrt{%
2/5}\pi ^{-3/4}a_{\perp }^{-3/2}g_{\mathrm{LHY}}$. Finally, the quartic LHY
term determines the asymptotic form of the wave functions at $r\rightarrow 0$
as $\sim r^{-2/3}$, which provides for the convergence of the 2D norm, i.e.,
it secures the existence of the GS in the 2D model including the LHY term.

\subsection{A challenging issue: the Fermi gas pulled to the center}

An interesting possibility is to elaborate the 3D model for the gas of
fermions pulled to the center by potential (\ref{U}). In a rigorous form,
this is a challenging problem, as the dynamical theory for Fermi gases
cannot be reduced to a simple mean-field equation \cite{Fermi}.
Nevertheless, there is a relatively simple approach to the description of
stationary states in a sufficiently dense gas, which is based on a
time-independent equation for the real fermionic wave function, $\Phi \left(
\mathbf{r}\right) $ \cite{Minguzzi,SKA,Luca}, with a nonlinear term of power
$7/3$ generated by the density-functional approximation, even in the absence
of direct interaction between the fermions, which is forbidden by the Pauli
principle. In the scaled form, this equation, including potential (\ref{U})
and chemical potential $\mu $, is%
\begin{equation}
\mu \Phi =-\frac{1}{3}\nabla ^{2}\Phi +\Phi ^{7/3}-\frac{U_{0}}{2r^{2}}\Phi .
\label{Phi}
\end{equation}%
The asymptotic form of the solution to Eq. (\ref{Phi}) at $r\rightarrow 0$ is%
\begin{equation}
\Phi (r)=\sqrt{\left( 3+4U_{0}\right) /8}r^{-3/2}+O\left( r^{1/2}\right) .
\label{Phi(r)}
\end{equation}%
This result demonstrates a problem similar to the one stressed above in the
case of the 2D model with the cubic nonlinearity: the substitution of
expression (\ref{Phi(r)}) in the 3D integral (\ref{N}) for the norm of the
wave function leads to the logarithmic divergence at $r\rightarrow 0$, hence
the relatively weak nonlinearity in Eq. (\ref{Phi}) is insufficient for the
suppression of the 3D quantum collapse of the Fermi gas pulled to the center
by potential (\ref{U}), and a more sophisticated analysis is necessary in
this case.

\section{The three-dimensional model with cylindrical symmetry}

The presentation in this section follows the original analysis reported in
Ref. \cite{HS2}.

\subsection{Formulation of the model}

The previous section addressed the most fundamental spherically symmetric
configuration in the 3D space. Because the geometry plays a crucially
important role in determining properties of the bound states produced by the
model, it is interesting to consider physically relevant settings in 3D with
the spatial symmetry reduced from spherical to a lower one. In particular,
it is possible to consider the model in which a strong uniform external
field is applied to the quantum gas, so that all the dipole moments, carried
by the particles, are polarized not towards the center, but in a fixed
direction ($z$), so that $\mathbf{d}=d\mathbf{e}_{z}$. This configuration
gives rise the cylindrically symmetric potential of the interaction of the
dipolar particle with the fixed attractive center:
\begin{equation}
U(\mathbf{r})=-\mathbf{d}\cdot \mathbf{E}_{Q}=-\frac{1}{2}U_{0}r^{-2}\cos
\theta ,  \label{cyl}
\end{equation}%
where $\cos \theta \equiv z/r$.

If the polarizing external field is electric, it acts on the central charge
too. For this reason, a more relevant situation corresponds to the case when
the ultracold gas is composed of \textit{Hund A }type of molecules, with
mutually locked electric and magnetic dipoles. Then, external uniform
magnetic field may be employed to align the dipoles in the fixed direction
\cite{Santos}.

The 3D GPE with potential (\ref{cyl}) is
\begin{equation}
i\frac{\partial \psi }{\partial t}=-\frac{1}{2}\left( \nabla ^{2}\psi +\frac{%
U_{0}}{r^{2}}\cos \theta \right) \psi +|\psi |^{2}\psi .  \label{nonlin}
\end{equation}%
Along with the consideration of the GS, it is also relevant to construct
eigenmodes carrying the orbital angular momentum, which corresponds to the
azimuthal quantum number, $m$:
\begin{equation}
\psi =e^{-i\mu t}e^{im\varphi }\Phi (r,\theta ),  \label{chiphi}
\end{equation}%
where the spherical coordinates are used again, cf. Eq. (\ref{psi}), and
real eigenmode $\phi $ should be found as a solution of the equation
following from the substitution of ansatz (\ref{chiphi}) in Eq. (\ref{nonlin}%
):
\begin{equation}
\mu \Phi =-\frac{1}{2}\left[ \frac{\partial ^{2}}{\partial r^{2}}+\frac{2}{r}%
\frac{\partial }{\partial r}-\frac{m^{2}}{r^{2}\sin ^{2}\theta }+\frac{1}{%
r^{2}\sin \theta }\frac{\partial }{\partial \theta }\left( \sin \theta \frac{%
\partial }{\partial \theta }\right) +\frac{U_{0}}{r^{2}}\cos \theta \right]
\Phi +\Phi ^{3}.  \label{chi-nonlin}
\end{equation}

\subsection{The linear Schr\"{o}dinger equation with the cylindrical symmetry%
}

The analysis of the present model should start with identifying conditions
for the existence of the GS in the respective linear Schr\"{o}dinger
equation, obtained by dropping the cubic term in Eq. (\ref{chi-nonlin}). At $%
r\rightarrow 0$, an asymptotic solution to the linear equation is looked for
as
\begin{equation}
\Phi \left( r,\theta \right) =r^{-\sigma }\chi _{\mathrm{lin}}(\theta ).
\label{sigma}
\end{equation}%
The substitution of ansatz (\ref{sigma}) in the linearized version of Eq. (%
\ref{chi-nonlin}) and dropping term $\mu \Phi $, which is negligible for the
asymptotic analysis at $r\rightarrow 0$, leads to an equation that can be
written in terms of $\xi \equiv \cos \theta $:
\begin{equation}
\frac{d}{d\xi }\left( \left( 1-\xi ^{2}\right) \frac{d\chi _{\mathrm{lin}}}{%
d\xi }\right) +\left( \sigma ^{2}-\sigma -\frac{m^{2}}{1-\xi ^{2}}+U_{0}\xi
\right) \chi _{\mathrm{lin}}(\xi )=0.  \label{xi}
\end{equation}%
For $U_{0}=0$, equation (\ref{xi}) with integer values
\begin{equation}
\sigma =l+1  \label{sigma-l}
\end{equation}%
may be solved in terms of the associated Legendre functions, $l\geq m$ being
the orbital quantum number. Note that the singular wave function (\ref{sigma}%
) is 3D-normalizable, at $r\rightarrow 0$, for $\sigma <3/2$, i.e., as it
follows from Eq. (\ref{sigma-l}), solely for the GS, with $m=l=0$ and $%
\sigma =1$.

\begin{figure}[t]
\begin{center}
\includegraphics[height=4.cm]{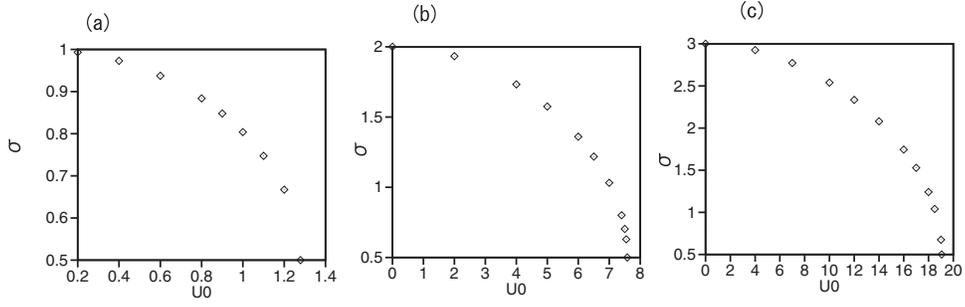}
\end{center}
\caption{ Eigenvalue $\protect\sigma $ of the singular eigenmode (\protect
\ref{sigma}), generated, as per Ref. \protect\cite{HS2}, by the numerical
solution of linear equation (\protect\ref{xi}), vs. strength $U_{0}$ of the
attractive potential, for three values of the azimuthal quantum number: (a) $%
m=0$, (b) $m=1$, (c) $m=2$. The eigenmode disappears, signaling the onset of
the quantum collapse, at $U_{0}>\left( U_{0}\right) _{\mathrm{cr}}(m)$, see
Eq. (\protect\ref{cr}), where $U_{0}=\left( U_{0}\right) _{\mathrm{cr}}(m)$
corresponds to $\protect\sigma =1/2$.}
\label{fig4}
\end{figure}

The onset of the quantum collapse is signalled by a transition in Eq. (\ref%
{xi}) from real eigenvalues $\sigma $ to complex ones. Because the effective
eigenvalue in the equation is $\epsilon \equiv \sigma ^{2}-\sigma $, i.e., $%
\sigma =(1+\sqrt{1+4\epsilon })/2$, the transition to complex $\sigma $
happens at point $\epsilon =-1/4$ (i.e., $\sigma =1/2$). At $U_{0}\neq 0$,
Eq. (\ref{xi}) cannot be solved in terms of standard special functions. A
result of a numerical solution is displayed in Fig. \ref{fig4}. It
demonstrates that, for given azimuthal quantum number $m$, eigenvalue $%
\sigma $ decreases, with the increase of $U_{0}$ from zero to some critical
value $\left( U_{0}\right) _{\mathrm{cr}}$, from $\sigma \left(
U_{0}=0\right) =m+1$ to $\sigma \left( U_{0}=\left( U_{0}\right) _{\mathrm{cr%
}}\right) =1/2$. For lowest values of $m$, the numerically found critical
values of the potential strength, at which $\sigma =1/2$ is attained are
\begin{equation}
\left( U_{0}\right) _{\mathrm{cr}}(m=0,1,2)=1.28,7.58,19.06.  \label{cr}
\end{equation}%
In the framework of the linear Schr\"{o}dinger equation, the quantum
collapse takes place at $U_{0}>\left( U_{0}\right) _{\mathrm{cr}}(m)$.
It is relevant to compare critical values (\ref{cr}) of the strength of the
axisymmetric potential with those given by Eq. (\ref{Ul}) for the
spherically isotropic one:

\begin{equation}
\left( U_{0}\right) _{\mathrm{cr}}^{\mathrm{(iso)}}(m=0,1,2)=\frac{1}{4}%
+m(m+1)\equiv 0.25,2.25,6.25.  \label{iso}
\end{equation}%
The comparison naturally shows that the critical strengths are much lower
for the spherical potential, which provides stronger pull to the center.

\subsection{ Suppression of the quantum collapse by the repulsive
nonlinearity under the cylindrical symmetry}

As well as in the isotropic setting, cf. Eq. (\ref{psichi3D}), the repulsive
cubic term in Eq. (\ref{chi-nonlin}) may balance the attractive potential $%
\sim -r^{-2}$ if, at $r\rightarrow 0$, the wave function contains the
singular factor $r^{-1}$ (rather than generic $r^{-\sigma }$ in Eq. (\ref%
{sigma})). Then, the substitution of%
\begin{equation}
\Phi (r,\theta )=r^{-1}\chi (r,\theta ),  \label{phi}
\end{equation}%
transforms Eq. (\ref{chi-nonlin}) into the following equation:
\begin{equation}
\mu \chi =-\frac{1}{2}\left[ \frac{\partial ^{2}\chi }{\partial r^{2}}+\frac{%
1-\xi ^{2}}{r^{2}}\frac{\partial ^{2}\chi }{\partial \xi ^{2}}-\frac{2\xi }{%
r^{2}}\frac{\partial \chi }{\partial \xi }+\left( U_{0}\xi -\frac{m^{2}}{%
1-\xi ^{2}}\right) \frac{\chi }{r^{2}}\right] +\frac{\chi ^{3}}{r^{2}}
\label{nonlin2}
\end{equation}%
(recall $\xi \equiv \cos \theta $). Note that Eq. (\ref{phi}) makes it
possible to write the norm of the 3D wave function as%
\begin{equation}
\frac{N}{2\pi }=\int_{0}^{\infty }r^{2}dr\int_{0}^{\pi }\sin \theta d\theta
\left\vert \psi \left( r,\theta \right) \right\vert ^{2}=\int_{0}^{\infty
}dr\int_{-1}^{+1}d\xi \chi ^{2}\left( r,\xi \right) .  \label{NN}
\end{equation}

To analyze solutions to Eq. (\ref{nonlin2}) at $r\rightarrow 0$, one may
expand them as%
\begin{equation}
\chi (r,\xi )=\chi _{0}(\xi )+\chi _{1}(\xi )r^{s/2},
\end{equation}%
assuming $s>0$, which leads to the following equation for $\chi _{0}(\xi )$,
that does not admit an exact solution:
\begin{equation}
\left( 1-\xi ^{2}\right) \frac{d^{2}\chi _{0}}{d\xi ^{2}}-2\xi \frac{d\chi
_{0}}{d\xi }+\left( U_{0}\xi -\frac{m^{2}}{1-\xi ^{2}}\right) \chi _{0}-\chi
_{0}^{3}=0,  \label{first}
\end{equation}%
cf. Eq. (\ref{xi}).

\begin{figure}[t]
\begin{center}
\includegraphics[height=5.cm]{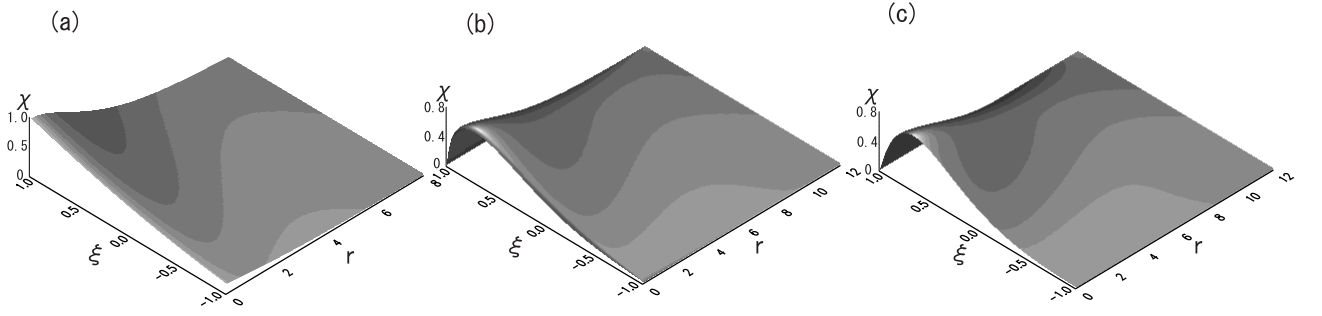}
\end{center}
\caption{Typical profiles of real function $\protect\chi (r,\protect\xi )$,
produced, in Ref. \protect\cite{HS2}, by the numerical solution of Eq.~(%
\protect\ref{nonlin2}), which determines the shape of the bound state with
the reduced (cylindrical) symmetry, as per Eq. (\protect\ref{phi}): (a) $%
m=0,U_{0}=3$; (b) $m=1,U_{0}=8.5$; (c) $m=2,U_{0}=20$. The solutions are
subject to normalization $N=2\protect\pi $, see Eq. (\protect\ref{NN}).}
\label{fig5}
\end{figure}

Bound states produced by Eq. (\ref{nonlin2}) were found by means of
numerical methods in Ref. \cite{HS2}. Typical profiles of solutions for
function $\chi \left( r,\xi \right) $, generated \ by Eq. (\ref{nonlin2}),
are displayed in Fig. \ref{fig5}, for $m=0,1,2$ and fixed norm, $N=2\pi $.

A crude analytical form of the solutions is provided by the Thomas-Fermi
approximation (TFA), which neglects all derivatives in Eq. (\ref{nonlin2})
\cite{BEC}:%
\begin{equation}
\chi _{\mathrm{TFA}}^{2}\left( r,\xi \right) =\left\{
\begin{array}{c}
\frac{1}{2}U_{0}\xi -\frac{m^{2}}{2\left( 1-\xi ^{2}\right) }-|\mu |r^{2},~~%
\mathrm{at}~~r^{2}<\frac{1}{2|\mu |}\left( U_{0}\xi -\frac{m^{2}}{1-\xi ^{2}}%
\right) , \\
0,~~\mathrm{at}~~r^{2}\geq \frac{1}{2|\mu |}\left( U_{0}\xi -\frac{m^{2}}{%
1-\xi ^{2}}\right) .%
\end{array}%
\right.  \label{Thomas}
\end{equation}%
Actually, this approximation for $m\geq 1$ exists only for $U_{0}>\left( 3%
\sqrt{3}/2\right) m^{2}$ (otherwise, Eq. (\ref{Thomas}) yields $\chi _{%
\mathrm{TF}}^{2}\equiv 0$).

\begin{figure}[t]
\begin{center}
\includegraphics[height=4.5cm]{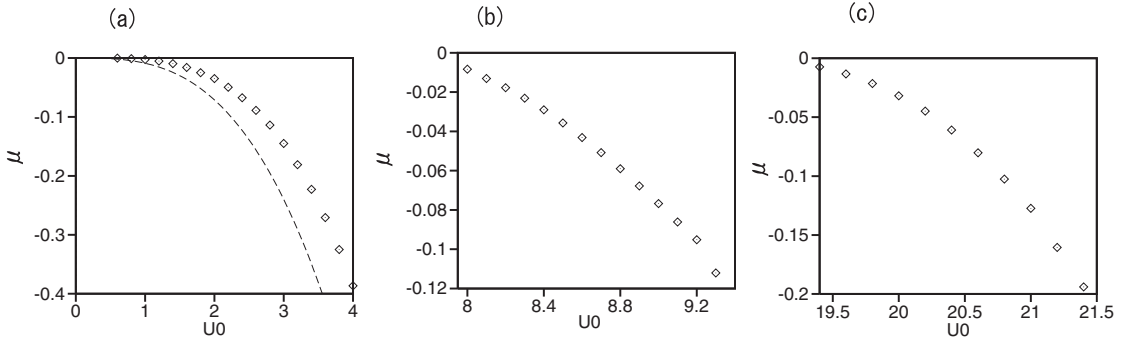}
\end{center}
\caption{Panels (a), (b), (c) display, respectively, the chemical potential
of the bound states with azimuthal quantum numbers $m=0,1,2$, vs. the
strength of the attractive potential, $U_{0}$, of the potential (\protect\ref%
{cyl}), with the the reduced (cylindrical) symmetry, and for the fixed norm,
$N=2\protect\pi $, as obtained in Ref. \protect\cite{HS2}. The dashed curve
in (a) additionally shows dependence (\protect\ref{muTF}) predicted by the
TF approximation.}
\label{fig6}
\end{figure}

Families of the bound states with different quantum numbers $m$ are
presented in Fig. \ref{fig6} by a set of curves showing the chemical
potential, $\mu $, versus nonlinearity strength, $U_{0}$, for a fixed norm ($%
N=N_{0}\equiv 2\pi $; producing the results for a fixed norm is sufficient,
as the scaling invariance of Eq. (\ref{nonlin2}) implies an exact property, $%
\mu \left( U_{0},N\right) =\left( N/N_{0}\right) ^{-2}\mu \left(
U_{0},N_{0}\right) $, the same as mentioned above for the isotropic
configuration). Figures \ref{fig6}(b) and (c) display the $\mu (U_{0}$)
dependences in relatively narrow intervals of values of $U_{0}$, to stress
that the dependences are obtained \emph{above} the critical values for the
linear Schr\"{o}dinger equation given by Eq. (\ref{cr}), where the linear
equation fails to produce any bound state.

TFA based on Eq. (\ref{Thomas}) makes it possible to predict the $\mu \left(
U_{0}\right) $ dependence for the GS ($m=0$) in an analytical form:
\begin{equation}
\mu _{\mathrm{TFA}}^{\mathrm{(GS)}}=-\left( 2/225\right) U_{0}^{3}.
\label{muTF}
\end{equation}%
As seen in Fig. \ref{fig6}(a), this simple approximation is reasonably close
to its numerically found counterpart.

Lastly, the stability of the bound states against perturbations was verified
in Ref. \cite{HS2} by direct simulations of the underlying GPE (\ref{nonlin}%
), demonstrating complete stability that the families of the states for $%
m=0,1,$ and $2$.

\section{The two-component system in three dimensions: the suppression of
the quantum collapse in miscible and immiscible settings}

This section summarizes results of the analysis reported in Ref. \cite{HS3}.

\subsection{The formulation of the model and analytical considerations}

The generalization of basic model (\ref{GPE2}) for a binary bosonic gas,
with component wave functions $\psi _{1}$ and $\psi _{2}$, is provided by
the system of nonlinearily coupled GPEs:
\begin{eqnarray}
i\frac{\partial \psi _{1}}{\partial t} &=&-\frac{1}{2}\nabla ^{2}\psi
_{1}+(|\psi _{1}|^{2}+\gamma |\psi _{2}|^{2})\psi _{1}-\frac{V_{0}}{r^{2}}%
\psi _{1}~,  \notag \\
&&  \label{GP} \\
i\frac{\partial \psi _{2}}{\partial t} &=&-\frac{1}{2}\nabla ^{2}\psi
_{2}+(\gamma |\psi _{1}|^{2}+|\psi _{2}|^{2})\psi _{2}-\frac{V_{0}}{r^{2}}%
\phi _{2}~  \notag
\end{eqnarray}%
(for the consistency with Ref. (\cite{HS2}), parameter $V_{0}\equiv U_{0}/2$
is now used as the strength of the potential pulling particles to the
center), where $\gamma $ is the relative strength of the inter-component
repulsion, and coefficients of the self-repulsion are scaled to be $1$.

Spherically symmetric bound states with chemical potentials $\mu _{n}<0$, $%
n=1,2$, of the two components are looked for as
\begin{equation}
\psi _{n}\left( r,t\right) =\frac{\chi _{n}(r)}{r}\exp \left( -i\mu
_{n}t\right) ,
\end{equation}%
with real radial functions $\chi _{n}(r)$ obeying the coupled equations,
\begin{eqnarray}
\mu _{1}\chi _{1} &=&-\frac{1}{2}\chi _{1}^{\prime \prime }-\frac{V_{0}}{%
r^{2}}\chi _{1}+\left( \chi _{1}^{2}+\gamma \chi _{2}^{2}\right) \frac{\chi
_{1}}{r^{2}},  \notag \\
&&  \label{stat} \\
\mu _{2}\chi _{2} &=&-\frac{1}{2}\chi _{2}^{\prime \prime }-\frac{V_{0}}{%
r^{2}}\chi _{2}+\left( \chi _{2}^{2}+\gamma \chi _{1}^{2}\right) \frac{\chi
_{2}}{r^{2}},  \notag
\end{eqnarray}%
cf. Eqs. (\ref{psichi3D}) and (\ref{chi}). In terms of these functions, the
norms of the components are%
\begin{equation}
N_{n}\equiv \int \left\vert \phi _{n}(\mathbf{r})\right\vert d\mathbf{r}%
=4\pi \int_{0}^{\infty }\left[ \chi _{n}(r)\right] ^{2}dr,  \label{NNN}
\end{equation}%
and the rms radial size of the trapped mode in each component is defined as%
\begin{equation}
\left\langle r_{n}^{2}\right\rangle =\frac{\int_{0}^{\infty }\left[ \chi
_{n}(r)\right] ^{2}r^{2}dr}{\int_{0}^{\infty }\left[ \chi _{n}(r)\right]
^{2}dr},  \label{<>}
\end{equation}%
cf. Eq. (\ref{R}).

An expansion of solutions to Eqs.~(\ref{stat}) at $r\rightarrow 0$ is looked
for as
\begin{equation}
\chi _{n}(r)=\chi _{n}^{(0)}\left[
1-c_{n}^{(1)}r^{s/2}-c_{n}^{(2)}r^{s/2+2}+\cdots
-d_{n}^{(1)}r^{2}-d_{n}^{(2)}r^{4}+\cdots \right] ,  \label{chi-n}
\end{equation}%
with $s>0$, cf. Eq. (\ref{r=0}) (here, $c_{1}\neq c_{2}$ is possible, but
power $s$ must be the same for $\chi _{1}$ and $\chi _{2}$), which leads to
a system of algebraic equations for leading-order coefficients $\chi
_{n}^{(0)}$:
\begin{eqnarray}
\chi _{1}^{(0)}\left[ \left( \chi _{1}^{(0)}\right) ^{2}+\gamma \left( \chi
_{2}^{(0)}\right) ^{2}\right] &=&V_{0}\chi _{1}^{(0)},  \notag \\
&&  \label{chichi} \\
\chi _{2}^{(0)}\left[ \left( \chi _{2}^{(0)}\right) ^{2}+\gamma \left( \chi
_{1}^{(0)}\right) ^{2}\right] &=&V_{0}\chi _{2}^{(0)}.  \notag
\end{eqnarray}%
Equations (\ref{chichi}) give rise to solutions of two types, corresponding
to mixed and demixed states in the binary gas:%
\begin{eqnarray}
\chi _{1}^{(0)} &=&\chi _{2}^{(0)}\equiv \chi _{\mathrm{mix}}^{(0)}=\sqrt{%
V_{0}/\left( 1+\gamma \right) };  \label{mix} \\
\chi _{1}^{(0)} &\equiv &\chi _{\mathrm{demix}}^{(0)}=\sqrt{V_{0}},~\chi
_{2}^{(0)}=0.  \label{demix}
\end{eqnarray}%
The numerical analysis performed in Ref. \cite{HS2} has demonstrates that
demixed modes do not exist at $\gamma <1$, when the mutual repulsion is
weaker than the self-repulsive nonlinearity, while mixed ones are completely
unstable in the opposite case, $\gamma >1$. Thus, unlike other systems
featuring the miscibility-immiscibility transitions \cite{misc0}-\cite{misc1}%
, in the present situation the transition point is not shifted, under the
action of the confining potential, from the commonly known free-space point,
$\gamma =1$ \cite{Mineev}.

Further analysis demonstrates a change in the structure of the $r$-dependent
corrections in Eq. (\ref{chi-n}) for the miscible system, with $\gamma <1$:
at $V_{0}<1/2$, the dominant terms are $\sim r^{\left( 1+\sqrt{1+16V_{0}}%
\right) /2}$, while at $V_{0}>1/2$ these are terms $\sim r^{2}$. This
\textit{break of analyticity,} which happens, with the increase of $V_{0}$,
at $V_{0}=1/2$, implies that a \textit{weak quantum phase transition }%
happens at this value of $V_{0}$, although the well-defined GS exists
equally well at $V_{0}<1/2$ and $V_{0}>1/2$. Precisely at $V_{0}=1/2\equiv
\left( V_{0}\right) _{\mathrm{phase-trans}}$, expansion (\ref{chi-n}) is
replaced by%
\begin{equation}
\chi _{n}(r)=\frac{1}{\sqrt{2\left( 1+\gamma \right) }}\left[ 1+\frac{\mu
_{1}+\mu _{2}}{4}r^{2}\ln \left( \frac{r_{0}}{r}\right) +(-1)^{n}\frac{%
1+\gamma }{4\gamma }\left( \mu _{1}-\mu _{2}\right) r^{2}\right] .
\label{ln}
\end{equation}

Note that the present phase transition is weak in comparison with the
above-mentioned one, driven by the LHY correction to the mean-field theory,
which gives rise to the jump between the different asymptotic forms of the
wave function, given by Eqs. (\ref{-2/3}) and (\ref{sigma-}). For the
comparison with the present setting, based on the binary BEC, especially
relevant are previously investigated phase transitions in binary fluids \cite%
{fluids}. Recall that the onset of the quantum collapse in the linear
version of the model occurs at the critical value$\left( V_{0}^{(\mathrm{cr}%
)}\right) _{1}\equiv (1/2)\left( U_{0}\right) _{\mathrm{cr}}^{(\mathrm{3D}%
)}=1/8$ \cite{LL}, which is a quarter of the value of the potential's
strength at the phase-transition point, $\left( V_{0}\right) _{\mathrm{%
phase-trans}}=1/2$.

Lastly, at $r\rightarrow \infty $ Eqs. (\ref{stat}) yield an exponential
asymptotic form of the solution,
\begin{equation}
\chi _{n}(r)\approx \chi _{n}^{(\infty )}\left( 1-\frac{V_{0}}{\sqrt{-2\mu
_{n}}r}\right) \exp \left( -\sqrt{-2\mu _{n}}r\right) ,
\end{equation}%
where constants $\chi _{n}^{(\infty )}$ are indefinite in terms of the
asymptotic expansion at $r\rightarrow \infty $.

\subsection{Numerical and additional analytical results for trapped binary
modes}

\subsubsection{Mixed ground states}

\begin{figure}[tbp]
\begin{center}
\includegraphics[height=4.cm]{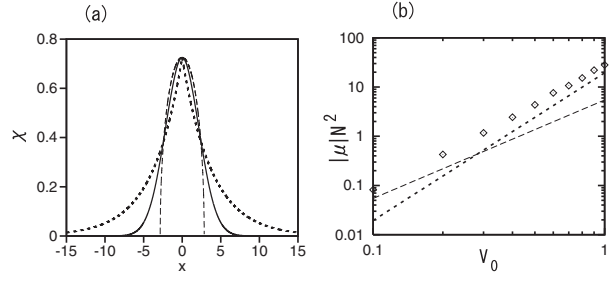}
\end{center}
\caption{(a) The numerically found profile of wave functions $\protect\chi %
_{1}(r)=\protect\chi _{2}(r)$, at $V_{0}=1$, $\protect\gamma =0.9$, and $%
N_{1}=N_{2}=4\protect\pi $, of the GS in the miscible binary system, as
found in Ref. \protect\cite{HS3}, and its comparison with the analytical
approximation given by Eq.~(\protect\ref{old}), and TFA based on Eq.~(%
\protect\ref{new}) (the short- and long-dashed lines, respectively). (b) The
chain of rhombuses depicts the numerically found relation between $|\protect%
\mu |N^{2}$ and $V_{0}$ at $\protect\gamma =0.9$. The short- and long-dashed
lines represent the approximations provided by Eqs. (\protect\ref{oldN}) and
(\protect\ref{e2}), respectively.}
\label{fig7}
\end{figure}

Figure \ref{fig7}(a) shows a typical profile for the mixed GS produced by a
numerical solution of Eq.~(\ref{stat}) at $V_{0}=1$ for $\gamma =0.9$ and
equal norms of the two components, $N_{1}=N_{2}=4\pi $.
The simplest global analytical approximation for the GS\ wave function is
provided by the interpolation, similar to that introduced in the
single-component setting, cf. Eq. (\ref{inter}):
\begin{equation}
\chi _{n}(r)\approx \chi _{\mathrm{mix}}^{(0)}e^{-\sqrt{-2\mu _{n}}r}.
\label{old}
\end{equation}%
The substitution of this interpolation in Eqs. (\ref{NNN}) and (\ref{<>}),
along with expression (\ref{mix}), leads to predictions for the chemical
potentials and squared average radius of the two components as functions of
their norms (which are valid too in the case of $N_{1}\neq N_{2}$):
\begin{eqnarray}
\mu _{n} &=&-2\left[ \frac{\pi V_{0}}{(1+\gamma )N_{n}}\right] ^{2},
\label{oldN} \\
\left\langle r_{n}^{2}\right\rangle &=&\left[ \frac{(1+\gamma )N_{n}}{2\pi
V_{0}}\right] ^{2}.  \label{oldr^2}
\end{eqnarray}%
Comparison of expression (\ref{oldN}) with numerical results is shown in
Fig.~\ref{fig7}(b) by the dashed line. This approximation is accurate for
sufficiently small $V_{0}$, but becomes inaccurate for large $V_{0}$.

For larger $V_{0}$, TFA can be applied to the mixed balanced mixture, with $%
N_{1}=N_{2}\equiv N$, which yields (for $\chi _{1}=\chi _{2}\equiv \chi $)
\begin{equation}
\chi _{\mathrm{TFA}}(r)=\left\{
\begin{array}{c}
\sqrt{\left( V_{0}+\mu r^{2}\right) /\left( 1+\gamma \right) }\,,~~\mathrm{at%
}~~r<R_{0}\equiv \sqrt{V_{0}/(-\mu )}, \\
0,~~\mathrm{at}~~r>R_{0}~,%
\end{array}%
\right.  \label{new}
\end{equation}%
cf. TFA for the potential with the cylindrical symmetry, given by Eq. (\ref%
{Thomas}). The substitution of approximation (\ref{new}) in Eqs. (\ref{N})
and (\ref{<>}) yields the predictions for the chemical potential and
effective size of the GS:
\begin{eqnarray}
\mu _{\mathrm{TFA}} &=&-\frac{64\pi ^{2}V_{0}^{3}}{9(1+\gamma )^{2}N^{2}},
\label{e2} \\
\left\langle r_{\mathrm{TFA}}^{2}\right\rangle &=&\frac{5}{\pi ^{3}}\left[
\frac{3\left( 1+\gamma \right) N}{16V_{0}}\right] ^{2}\equiv \frac{5}{4\pi }%
R_{0}^{2}  \label{e3}
\end{eqnarray}%
(recall $R_{0}$ is the TFA cutoff radius defined in Eq. (\ref{new})).
Analytical approximations (\ref{old}) and (\ref{new}) (shown by the short-
and long-dashed lines, respectively) are compared to the numerically found
profile of the GS in Fig. \ref{fig7}(b). A general conclusion (see details
in Ref. \cite{HS3}) is that, quite naturally, TFA works better for larger $%
V_{0}$, while interpolation (\ref{old}) is more accurate for smaller $V_{0}$.

Numerically generated profiles of imbalanced mixed GSs are displayed in Fig.~%
\ref{fig3}(a) at $V_{0}=2$ and $\gamma =0.9$ for $N_{1}=4\pi $ and $%
N_{2}=2\pi $. The imbalanced mixed states with $\mu _{1}\neq \mu _{2}$ and $%
N_{1}\neq N_{2}$ feature equal values of $\chi _{1,2}(r=0)$, in agreement
with Eq. (\ref{mix}).

\begin{figure}[tbp]
\begin{center}
\includegraphics[height=4.cm]{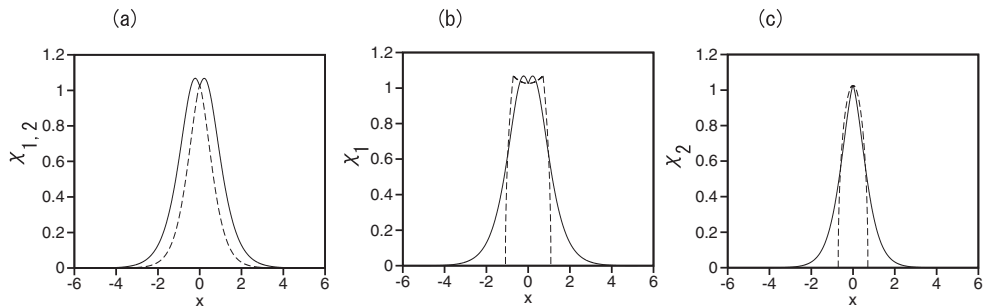}
\end{center}
\caption{(a) $\protect\chi _{1}$ (continuous) and $\protect\chi _{2}$
(dashed) components of the imbalanced mixed GS of the binary system, at $%
V_{0}=2$ and $\protect\gamma =0.9$, with $N_{1}=4\protect\pi $ and $N_{2}=2%
\protect\pi $, as found in Ref. \protect\cite{HS3}. (b) and (c): Comparison
of the numerical result (continuous lines) with the two-layer TFA (dashed
lines, see Eqs. (\protect\ref{TF-two}) and (\protect\ref{out})) for $\protect%
\chi _{1}(r)$ and $\protect\chi _{2}(r)$. }
\label{fig8}
\end{figure}

In the case of the strong pull to the center, $V_{0}\gg 1$, TFA can be
generalized for imbalanced states, fixing $\left\vert \mu _{1}\right\vert
\leq \left\vert \mu _{2}\right\vert $ for the definiteness' sake. Then, TFA
is constructed in a two-layer form, technically similar to that applied to
the so-called symbiotic gap solitons in Ref. \cite{Thawatchai}. In the
\textit{inner layer},
\begin{equation}
r^{2}<r_{0}^{2}\equiv \frac{1-\gamma }{\gamma \mu _{1}-\mu _{2}}V_{0},
\label{r0}
\end{equation}%
both wave functions are different from zero:%
\begin{equation}
\chi _{n}^{\mathrm{(inner)}}(r)=\sqrt{\frac{V_{0}}{1+\gamma }-\frac{\gamma
\mu _{3-n}-\mu _{n}}{1-\gamma ^{2}}r^{2}}.  \label{TF-two}
\end{equation}%
In the \textit{outer layer}, only one component is present, in the framework
of TFA: $\chi _{2}\equiv 0$,%
\begin{equation}
\chi _{1}^{\mathrm{(outer)}}(r)=\left\{
\begin{array}{c}
\sqrt{V_{0}+\mu _{1}r^{2}},~\mathrm{at}~~r_{0}^{2}\leq r^{2}\leq
R_{0}^{2}\equiv -V_{0}/\mu _{1}, \\
0,~\mathrm{at}~~r^{2}\geq R_{0}^{2}~.%
\end{array}%
\right.  \label{out}
\end{equation}%
Both components of the TFA solution, given by Eqs. (\ref{r0})-(\ref{out}),
are continuous at $r=r_{0}$ and $r=R_{0}$. The two-layer TFA for a typical
imbalanced GS is compared to its numerical counterpart in Figs. \ref{fig8}%
(b,c).

The analysis reported in Ref. \cite{HS3} also includes the consideration of
a two-component system with \emph{attraction} between the components, in the
case when only one component is subject to the action of the
pull-to-the-center potential, while the other one plays the role of a
buffer. In particular, the interpolation, similar to that based on Eq. (\ref%
{old}), produces a sufficiently accurate prediction in that case.

\subsubsection{Immiscible ground states}

\begin{figure}[tbp]
\begin{center}
\includegraphics[height=4.cm]{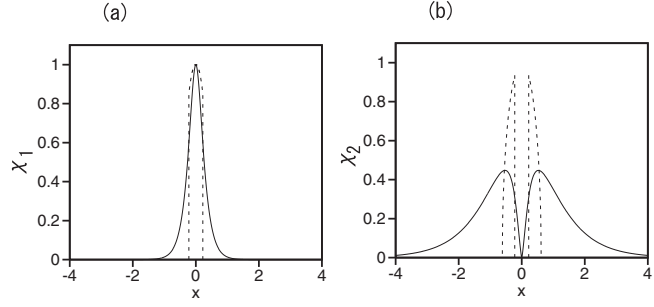}
\end{center}
\caption{(a) and (b): Comparison of the numerically found profiles for
components $\protect\chi _{1}(r)$ and $\protect\chi _{2}(r)$ of the
immiscible GS (solid lines) in the binary condensate ($V_{0}=1$, $\protect%
\gamma =1.2$) with equal norms of both components ($N_{1}=N_{2}=0.8\protect%
\pi $), and the respective TFA, given by Eqs. (\protect\ref{ImmTFin}) and (%
\protect\ref{ImmTFout}), respectively (continuous lines), as per Ref.
\protect\cite{HS3}. The numerical solution gives widely different values of
chemical potentials of the two components in this case: $\protect\mu %
_{1}=-14.2$, $\protect\mu _{2}=-0.84$.}
\label{fig9}
\end{figure}

As said above, in the case of $\gamma >1$ relevant states are immiscible
ones. The two-layer TFA may be applied to produce an immiscible GS. In the
inner layer,
\begin{equation}
r^{2}<r_{0}^{2}=\frac{(\gamma -1)V_{0}}{\gamma \mu _{1}-\mu _{2}},  \notag
\end{equation}%
the approximation yields%
\begin{equation}
\chi _{1}(r)=\sqrt{V_{0}+\mu _{1}r^{2}},\;\chi _{2}(r)=0.  \label{ImmTFin}
\end{equation}%
In the outer layer, which is $r_{0}^{2}<r^{2}<R_{0}^{2}=V_{0}/(-\mu _{2})$,
the result is
\begin{equation}
\chi _{1}(r)=0,\;\chi _{2}(r)=\sqrt{V_{0}+\mu _{2}r^{2}},  \label{ImmTFout}
\end{equation}%
i.e., TFA predicts complete separation between the components in the
immiscible state. Figure \ref{fig9} compares the approximation to numerical
results. Of course, the immiscible components are not completely separated
in the numerical solution.

\section{The mean-field predictions versus the many-body quantum theory}

\subsection{Introduction to the section}

The analysis presented above was performed in Refs. \cite{HS1}-\cite{HS3} in
the framework of the mean-field theory, i.e., the respective GPEs (possibly
including the beyond-mean-field LHY corrections, see Eq. (\ref{+LHY})). A
relevant issue is comparison of the basic mean-field predictions, such as
the suppression of the quantum collapse and creation of the originally
missing GS, with the consideration of the many-body system of repulsively
interacting quantum bosons, pulled to the center by potential (\ref{U}),
which is taken here as $U(r)=-U_{0}/r^{2}$, i.e., $U_{0}$ in Eq. (\ref{U})
is replaced by $2U_{0}$, to make the notation consistent with that in Ref.
\cite{GEA}. This problem was addressed in Ref. \cite{GEA}. Results produced
in that work are recapitulated in the present section.

The many-body Hamiltonian representing the setting under the consideration
is
\begin{equation}
\hat{H}=-\sum\limits_{j=1}^{N}\left( \frac{\hbar ^{2}\nabla _{j}^{2}}{2m}+%
\frac{U_{0}}{r_{j}^{2}}\right) +\sum\limits_{j<k}^{N}V_{\mathrm{int}}(|%
\mathbf{r}_{j}-\mathbf{r}_{k}|),  \label{H}
\end{equation}%
where $\mathbf{r}_{j}$ are coordinates of the $j$-th particle in the 3D
space, $m$ is the particle's mass, and $V_{\mathrm{int}}(r)$ is the
potential of the repulsive interaction between the particles. In the
framework of the mean-field theory, $V_{\mathrm{int}}(r)$ is characterized
solely by the $s$-wave scattering length~\cite{BEC}, while the many-body
system should be introduced with a particular form of the interaction
potential. Two basic forms of the interaction potential chosen for the
analysis are specified below, see Eqs. (\ref{Eq:V:HS}) and (\ref{Eq:V:soft}).

Before introducing the many-body wave function, the single-particle one is
adopted as per the following ansatz:
\begin{equation}
f_{1}(r)=r^{\beta }\exp (-\alpha r^{2}),  \label{Eq:f1}
\end{equation}%
where $\alpha \geq 0$ determines the inverse localization length, which
affects the system's size and, consequently, the density. Alternatively, $%
\alpha $ can be interpreted in terms of an effective external harmonic
confinement with frequency $\Omega =2\alpha \hbar /m$, cf. Eq. (\ref{trap}).
At $r_{j}\rightarrow 0$, the shape of the wave function is controlled by
parameter $\beta $ in ansatz~(\ref{Eq:f1}).

\subsection{The single-particle solution}

The single-particle problem, defined by Hamiltonian~(\ref{H}) with $N=1$,
can be studied by means of the variational method, treating $\alpha $ and $%
\beta $ in ansatz~(\ref{Eq:f1}) as variational parameters. In the
single-particle sector, the system is steered by the competition of the
external potential and kinetic energy, while the interparticle potential, $%
V_{\mathrm{int}}(|\mathbf{r}_{i}-\mathbf{r}_{j}|)$, does not appear. The
variational energy, $E_{\mathrm{var}}^{(1)}=\left[ \int f_{1}^{2}(\mathbf{r}%
)\;d\mathbf{r}\right] ^{-1}\int f_{1}(\mathbf{r})Hf_{1}(\mathbf{r})d\mathbf{r%
}$, with $f_{1}$ taken as per Eq.~(\ref{Eq:f1}), is
\begin{equation}
E^{(1)}=\alpha \left[ 1-\frac{8U_{0}-1}{2(1+2\beta )}\right] .  \label{Eq:E1}
\end{equation}%
For a fixed localization size, $\alpha =\mathrm{const}$, this energy is a
decreasing function of $\beta $ if $U_{0}$ is smaller than the critical
value for the onset of the collapse, $U_{0}=1/8$ -- the same which appears
in Eq. (\ref{1/4}). On the other hand, a \textit{metastable} state may
appear in the many-body system with repulsive interparticle interactions.
Actually, it corresponds to the mean-field GS predicted by the solution of
the GPE~in Ref. \cite{HS1}, see further details below

In the framework of the local-density approximation, the chemical potential
of the state with uniform density $n$ is take as $\mu _{\mathrm{\hom }}=gn$
, where $g=4\pi \hbar ^{2}a_{s}/m$ is the coupling constant. This choice
corresponds to the short-range interaction potential determined by the $s$%
-wave scattering length as per the Born approximation. Further, the chemical
potential in the presence of the external field is approximated by the sum
of the local chemical potential $\mu _{\mathrm{loc}}=gn$, where, this time, $%
n$ is a function of the coordinates, rather than a constant, and the
external potential,
\begin{equation}
\mu =\mu _{\mathrm{loc}}-\frac{U_{0}}{r^{2}}+\frac{1}{2}m\Omega ^{2}r^{2},
\label{Eq:LDA}
\end{equation}%
where the harmonic-oscillator confinement, with the respective length scale,
$a_{\mathrm{ho}}=\sqrt{\hbar /(m\Omega )}$, is added to make the size of the
system finite, cf. potential (\ref{trap}) used above. Solving Eq.~({\ref%
{Eq:LDA}}) for the density, one obtains the following density profile:
\begin{equation}
n(r)=\frac{1}{g}%
\begin{cases}
\mu -\frac{1}{2}m\Omega ^{2}r^{2}+U_{0}r^{-2}, & \mathrm{at\ }r<R_{\mathrm{%
TFA}}, \\
0, & \mathrm{at\ }\geq R_{\mathrm{TFA}},%
\end{cases}
\label{Eq:LDA:n}
\end{equation}%
where the radius of the gaseous cloud is taken as per TFA, $R_{\mathrm{TFA}}=%
\sqrt{\mu +\sqrt{\mu ^{2}+2mU_{0}\Omega ^{2}}}/(\sqrt{m}\Omega )$. The
density at the center features an integrable divergence in Eq. (\ref%
{Eq:LDA:n}), reflecting the presence of the attractive central potential,
cf. Eq. (\ref{psichi3D}). Finally, the chemical potential itself is fixed by
the normalization condition, $4\pi \int_{0}^{R_{\mathrm{TFA}}}n(r)r^{2}dr=N$.

To study the expected scenarios of the system's evolution, two different
potentials of the inter-particle interaction were introduced in Ref. \cite%
{GEA}, \textit{viz}., the hard-sphere potential of diameter $R$,
\begin{equation}
V_{\text{\textrm{hard}}}(r)=%
\begin{cases}
\infty , & r<R \\
0, & r\geq R%
\end{cases}%
,  \label{Eq:V:HS}
\end{equation}%
and its soft-sphere counterpart,
\begin{equation}
V_{\mathrm{soft}}(r)=%
\begin{cases}
V_{0}, & r<R \\
0, & r\geq R%
\end{cases}%
,  \label{Eq:V:soft}
\end{equation}%
with finite $V_{0}$ in the latter case. By varying height $V_{0}$ of the
soft-sphere potential, one can alter the respective $s$-wave scattering
length, which is
\begin{equation}
a_{s}=R[1-\tanh (kR)/(kR)],  \label{Eq:a:SS}
\end{equation}%
where the momentum corresponding to the height of the soft-sphere potential
is%
\begin{equation}
k\equiv \sqrt{mV_{0}}/\hbar ~.  \label{k}
\end{equation}%
For hard-sphere potential~(\ref{Eq:V:HS}), the effective $s$-wave scattering
length is identical to the diameter of the sphere, $a_{s}=R$.

\subsection{The Monte-Carlo method}

An efficient way to calculate the energy of a many-body system is to use the
Monte-Carlo technique. In Ref. \cite{GEA}, the variational Monte-Carlo
method was employed, which samples the probability distribution, $p=|\psi
|^{2}$, for a known many-body wave function, $\psi $, allowing one to
calculate the variational energy as a function of trial parameters, such as $%
\alpha $ and $\beta $ in Eq.~(\ref{Eq:f1}). The well-known Metropolis
algorithm \cite{Metropolis} was used for the implementation of the method.

The many-body trial wave function was chosen as a product of single-particle
terms, $f_{1}(r)$, taken as per Eq. (\ref{Eq:f1}), and a pairwise product of
two-particle Jastrow terms \cite{Jastrow}, $f_{2}(r)$:
\begin{equation}
\psi (\mathbf{r}_{1},\dots ,\mathbf{r}_{N})=\prod%
\limits_{j=1}^{N}f_{1}(r_{j})\prod\limits_{j<k}^{N}f_{2}(|\mathbf{r}_{j}-%
\mathbf{r}_{k}|)  \label{Eq:f2}
\end{equation}%
The Jastrow factor $f_{2}(r)$ in Eq.~(\ref{Eq:f2}) is chosen as a solution
of the linear Schr\"{o}dinger equation for two-body scattering. In this way,
interparticle correlations, which are important in the context of the
metastability of the many-body system, are retained in the analysis.

For the hard-sphere potential, the two-body solution is given by
\begin{equation}
f_{2}^{\mathrm{(hard)}}(r)=%
\begin{cases}
0, & r<R \\
1-R/r, & r\geq R%
\end{cases}%
,  \label{Eq:f2:HS}
\end{equation}%
while for the soft-sphere potential~(\ref{Eq:V:soft}), it is \cite{LL}
\begin{equation}
f_{2}^{\mathrm{(soft)}}(r)=%
\begin{cases}
A\sinh (kr)/r, & r<R \\
1-a_{s}/r, & r\geq R%
\end{cases}%
,  \label{Eq:f2:SS}
\end{equation}%
where $k$ is given by Eq. (\ref{k}), and constant $A$ is determined by the
condition of the continuity of $f_{2}(r)$ at $r=R$.

\subsection{Numerical results for the many-body system}

\begin{figure}[tbp]
\includegraphics[width=0.4\columnwidth, angle=0]{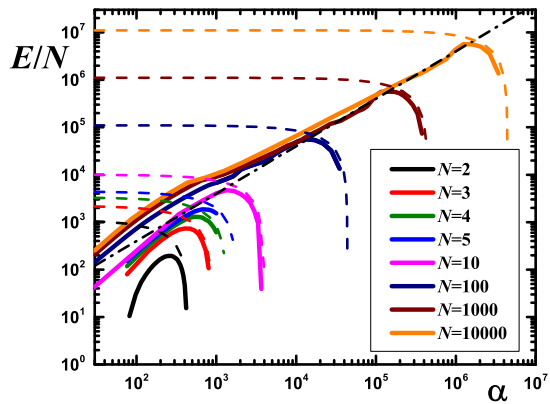}
\caption{(Color online) The energy per particle in the many-body system for
the soft-sphere interaction potential, as a function of the
inverse-Gaussian-width parameter, $\protect\alpha $ (see Eq. (\protect\ref%
{Eq:f1})), for $U_{0}=1$, $a_{s}=0.1$, $R=1.3a_{s}$ and the number of
particles $N=2,3,4,5,10,100,1000,10000$ (larger numbers of particles have a
larger value at the maximum), as obtained in Ref. \protect\cite{GEA}. Solid
lines: the variational result; dashed lines: the asymptotic energy of the
fully-collapsed state, as per Eq.~(\protect\ref{Eq:EN}); the dash-dotted
line: typical energy associated with the Gaussian localization, as given by
Eq.~(\protect\ref{Eq:E:C}). }
\label{fig10}
\end{figure}

Figure~\ref{fig10} shows the variational energy, calculated by the Monte
Carlo method for a fixed radius of the soft sphere, $R=1.3a_{s}$, and a wide
range of values of the number of particles, from $N=2$ up to $N=10000$. For
small values of $\alpha $, which corresponds to the weak localization, the
energy may be negative (this is not visible in the log-log plot of Fig.~\ref%
{fig10}). As the localization gets tighter, the energy becomes positive, as
the two-body interaction helps the system to resist the trend to collapsing.
For very tight localization, $\alpha \rightarrow \infty $, the collapse is
observed for small values of $N$, with the energy diverging towards $-\infty
$.

For large $N$, the energy calculated with ansatz~(\ref{Eq:f1}) does not
immediately lead to the fully collapsed state. The localization energy,
proportional to the energy scale, $\hbar \Omega $, of the trapping
potential, may become a dominating term in the energy, while the system's
size is still large enough, so that the fully-collapsed state is not
realized. The energy in the corresponding regime is numerically approximated
as
\begin{equation}
E=NC\alpha  \label{Eq:E:C}
\end{equation}%
with $C=4$. It is shown in Fig.~\ref{fig10} by the dashed-dotted line. For
still tighter localization, it has been found that, in the limit of the full
collapse, when all particles overlap, the energy is well approximated by
formula
\begin{equation}
E=NE^{(1)}+N(N-1)V_{0}.  \label{Eq:EN}
\end{equation}%
The energy of the interparticle interactions, revealed by the calculations,
is $E_{\mathrm{int}}=V_{0}N(N-1)/2$. The asymptotic energy~(\ref{Eq:EN}) is
shown in Fig.~\ref{fig3} by dashed lines.

A clear conclusion is that the increase of the number of particles indeed
causes strong rise of the potential barrier which stabilizes the metastable
energy minimum corresponding to the gaseous state. This can be also
concluded from Eq.~(\ref{Eq:EN}), where the contribution due to the
repulsive interactions scales as $N^{2}$ for large $N$, while the term
corresponding the attractive central potential scales as $N$.

The energy barrier between the state described by ansatz~(\ref{Eq:f1}) and
the free state with zero energy, $E_{\mathrm{barrier}}$, is estimated as the
maximum value of the energy per particle (see Fig.~\ref{fig3}), and is shown
in Fig.~\ref{fig4}. For a large system's size, the barrier can be
approximated by comparing the two basic energy scales given by Eqs.~(\ref%
{Eq:E:C}) and~(\ref{Eq:EN}). The resulting asymptotic approximation for the
barrier's height is
\begin{equation}
E_{\mathrm{barrier}}=6NV_{0}/(8U_{0}-3+2C),  \label{large}
\end{equation}%
which is shown in Fig.~\ref{fig4} by the dashed line.

\begin{figure}[tbp]
\includegraphics[width=0.4\columnwidth, angle=0]{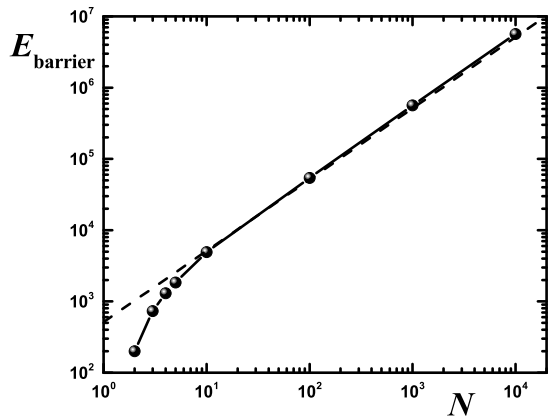}
\caption{The energy barrier between the state with $\protect\alpha =0$ and $%
\protect\alpha \rightarrow \infty $ in the many-body system, as a function
of the number of particles, $N$, for the data shown in Fig.~\protect\ref%
{fig10}, as per Ref. \protect\cite{GEA}. The dashed line depicts the
asymptotic approximation~(\protect\ref{large}) for the large system. }
\label{fig11}
\end{figure}

\section{Discussion and conclusion}

This article aims to produce a brief review of results reported in works
\cite{HS1}-\cite{HS3} and \cite{GEA}, that offer a solution to the known
problem of the quantum collapse, alias \textquotedblleft fall onto the
center" \cite{LL}, in nonrelativistic quantum mechanics. The quantum
collapse occurs in the three-dimensional Schr\"{o}dinger equation with 3D
isotropic attractive potential $-U_{0}/(2r^{2})$. This equation does not
have a GS (ground state) if the attraction strength, $U_{0}$, exceeds a
final critical value. In that case, the Schr\"{o}dinger equation gives rise
to a nonstationary wave function which collapses, shrinking towards the
center. The solution of the collapse problem was proposed in the
above-mentioned original works in terms of the gas of bosons pulled to the
center by the same potential, with repulsive contact interactions between
the particles. The intrinsic repulsion is represented by the cubic term in
the respective GPE (Gross-Pitaevskii equation). The setting may be realized
as the 3D gas of polar molecules carrying a permanent electric dipole moment
and pulled to a central electric charge. The analysis, performed in the
framework of the mean-field theory, predicts suppression of the collapse in
the gas, and the creation of the missing GS.

An original result, added in this article to the review of the previously
published findings, is the quantum phase transition occurring in the 3D
model which includes the beyond-mean-field LHY (Lee-Huang-Yang) correction
in the GPE, in the form of the self-repulsive quartic term. The phase
transition manifests itself by a jump of the asymptotic structure of the
wave function (for $r\rightarrow 0$) at the critical value of the strength
of the attractive potential.

In the 2D version of the model, the cubic self-repulsion is not sufficient
to suppress the quantum collapse. In this case, it can be suppressed if a
quintic self-repulsive term, representing three-body collisions (provided
that they so not give rise to losses) can be added to the underlying GPE. On
the other hand, the LHY quartic term, added to the 2D GPE, is sufficient to
suppress the quantum collapse and restore the respective GS.

Polarization of dipole moments in the 3D gas by an external uniform field
reduces the symmetry of the central attractive potential from spherical to
cylindrical. This modification of the system predicts both the GS and
stabilized states carrying the angular momentum. A binary condensate,
modelled by the system of nonlinearly-coupled GPEs, is considered too,
making it possible to study the interplay of the suppression of the collapse
in the 3D space and the miscibility-immiscibility transition in the binary
BEC.

In addition to the systematic numerical analysis of these mean-field
settings, the original works have produced many results by means of
analytical approximations, such as combined asymptotic expansions and TFA
(Thomas-Fermi approximation). All the states predicted by the mean-field
theory in these settings are shown to be completely stable as solutions to
the respective time-depending GPEs.

In work \cite{GEA}, the consideration of the same 3D setting was performed
in terms of the many-body quantum theory, by means of the variational
approximation for the many-body wave function, numerically handled with the
help of the Monte-Carlo method. The analysis has demonstrated that, although
the quantum collapse cannot be fully suppressed in terms of the many-body
theory, the self-trapped states predicted by the mean-field model exist in
the full many-body setting too, as metastable ones, protected against the
onset of the collapse by a tall potential barrier, whose height steeply
grows with the increase of the number of particles in the gas.

As an extension of the work on the topic of this article, it may be
interesting to construct modes carrying the angular momentum in the
isotropic 3D model, and also to consider the model with a set of two
mutually symmetric attractive centers. In particular, it may be relevant to
explore a possibility of the spontaneous symmetry breaking of the GS in the
latter case.

As mentioned above (see Eqs. (\ref{Phi}) and (\ref{Phi(r)}), a challenging
issue is to develop a consistent analysis for the gas of fermions pulled to
the center by the same potential, $\ -U_{0}/\left( 2r^{2}\right) $.

\section*{Acknowledgments}

I appreciate valuable collaborations with Hidetsugu Sakaguchi and Gregory
Astrakharchik, who were my coauthors in works \cite{HS1}-\cite{HS3} and \cite%
{GEA}, on which the mini-review is based.

My recent work on topics relevant to the mini-review is supported by the
joint program in physics between NSF and Binational (US-Israel) Science
Foundation through project No. 2015616, and by the Israel Science Foundation
through Grant No. 1286/17.


\section*{Conflict of interests}

The author declares no conflict of interest in the context of this paper.


\begin{thebibliography}{99}
\bibitem{LL} Landau, L. D.; Lifshitz, E. M. \textit{Quantum Mechanics:
Nonrelativistic Theory}. Nauka publishers: Moscow, USSR, 1974.

\bibitem{anomaly} Gupta, K. S.; Rajeev, S. G. Renormalization in quantum
mechanics. \emph{Phys. Rev. D} \textbf{1993}, \emph{48}, 5940-5945.

\bibitem{anomaly2} Camblong, H. E.; Epele, L. N.; Fanchiotti, H.; Garc\'{\i}%
a Canal, C. A. Renormalization of the Inverse Square Potential. \emph{Phys.
Rev. Lett}. \textbf{2000}, \emph{85}, 1590.

\bibitem{superselection} \'{A}vila-Aoki, M.;\textit{\ }Cisneros C.; Mart%
\'{\i}nez-y-Romero, R. P.; N\'{u}\~{n}ez-Yepez, H. N.; Salas-Brito, A. L.
Classical and quantum motion in an inverse square potential. \emph{Phys.
Lett. A} \textbf{373}, 418-421 (2009).

\bibitem{HS1} Sakaguchi, H.; Malomed, B. A. Suppression of the
quantum-mechanical collapse by repulsive interactions in a quantum gas.
\emph{Phys. Rev. A} \textbf{2011}, \emph{83}, 013607.

\bibitem{HS2} Sakaguchi, H.; Malomed, B. A. Suppression of the quantum
collapse in an anisotropic gas of dipolar bosons. \emph{Phys. Rev. A}
\textbf{2011}, \emph{84}, 033616.

\bibitem{HS3} Sakaguchi, H.; Malomed, B. A. Suppression of the quantum
collapse in binary bosonic gases. \emph{Phys. Rev. A} \textbf{2013}, \emph{88%
}, 043638).

\bibitem{BEC} Pitaevskii, L. and Stringari, S. \textit{Bose-Einstein
Condensation}. Clarendon: Oxford, UK, 2003.

\bibitem{ion} Schmid, S.; H\"{a}rter, A.; Denschlag, J. H. Dynamics of a
cold trapped Ion in a Bose-Einstein condensate. \emph{Phys. Rev. Lett.}
\textbf{2010}, \emph{105}, 133202.

\bibitem{LiCs} Deiglmayr, J.; Grochola, A.; Repp, M.; M\"{o}rtlbauer, K.; Gl%
\"{u}ck, C.; Lange, J.; Dulieu, O.; Wester, R.; Weidem\"{u}ller M. Formation
of ultracold polar molecules in the rovibrational ground state. \emph{Phys.
Rev. Lett.} \textbf{2008}, 101, 133004.

\bibitem{KRb} Ospelkaus, S.; Ni, K.-K.; Qu\'{e}m\'{e}ner, G.; Neyenhuis, B.;
Wang, D.; de Miranda, M. H. G.; Bohn, J. L.; Ye, J.; Jin, D. S. Controlling
the hyperfine state of rovibronic ground-state polar molecules. \emph{Phys.
Rev. Lett.} \textbf{2010}, \emph{104}, 030402.

\bibitem{2D-review} Posazhennikova, A. Colloquium: Weakly interacting,
dilute Bose gases in 2D. \emph{Rev. Mod. Phys.} \textbf{2006}, \emph{78},
1111-1134.

\bibitem{Schmiedmayer} Denschlag, J.; Schmiedmayer, J. Scattering a neutral
atom from a charged wire. \emph{Europhys. Lett}. \textbf{1997}, \emph{38},
405-410.

\bibitem{Olshanii} Olshanii, M.; Perrin, H.; Lorent, V. Example of a quantum
anomaly in the physics of ultracold gases. \emph{Phys. Rev. Lett.} \textbf{%
2010}, \emph{105}, 095302.

\bibitem{anti} Sakaguchi, H.; Malomed, B. A. Solitons in combined linear and
nonlinear lattice potentials. \emph{Phys. Rev. A} \textbf{2010}, \emph{81},
013624.

\bibitem{VK} Vakhitov, M.; Kolokolov, A. Stationary solutions of the wave
equation in a medium with nonlinearity saturation. \emph{Radiophys. Quantum
Electron}. \textbf{1973}, \emph{16}, 783-789.

\bibitem{Berge'} Berg\'{e}, L. Wave collapse in physics: principles and
applications to light and plasma waves. \emph{Phys. Rep}. \textbf{1998},
\emph{303}, 259-370.

\bibitem{Dodd} Dodd, R. J. Approximate solutions of the nonlinear Schr\"{o}%
dinger equation for ground and excited states of Bose-Einstein condensates.
\emph{J. Res. Natl. Inst. Stand. Technol}. \textbf{1996}, \emph{101},
545-552.

\bibitem{Stringari} Dalfovo, F., Stringari, S. Bosons in anisotropic traps:
Ground state and vortices. \emph{Phys. Rev. A} \textbf{1996}, \emph{53},
2477-2485.

\bibitem{Alexander} Alexander, T. J., Berg\'{e}, L. Ground states and
vortices of matter-wave condensates and optical guided waves. \emph{Phys.
Rev. E} \textbf{2002}, \emph{65}, 026611.

\bibitem{DumDum} Malomed, B. A., Lederer, F., Mazilu, D., Mihalache, D. On
stability of vortices in three-dimensional self-attractive Bose-Einstein
condensates. \emph{Phys. Lett. A} \textbf{2007}, \emph{361}, 336-340.

\bibitem{LHY} Lee, T. D.; Huang, K.; Yang, C. N. Eigenvalues and
eigenfunctions of a Bose system of hard spheres and its low-temperature
properties. \emph{Phys. Rev}. \textbf{1957}, \emph{106}, 1135-1145.

\bibitem{Petrov} Petrov, D. S. Quantum mechanical stabilization of a
collapsing Bose-Bose mixture.\emph{\ Phys. Rev. Lett}. \textbf{2015}, \emph{%
115}, 155302.

\bibitem{AP} Petrov, D. S.; Astrakharchik, G. E. Ultradilute low-dimensional
liquids, \emph{Phys. Rev. Lett.} \textbf{2016}, \emph{117}, 100401.

\bibitem{Grisha} Astrakharchik, G. E.; Gangardt, D. M.; Lozovik, Yu. E.;
Sorokin, I. A. Off-diagonal correlations of the Calogero-Sutherland model,
\emph{Phys. Rev. E} \textbf{2006}, \emph{74}, 021105.

\bibitem{other1} Chubukov, A. V.; P\'{e}pin, C.; Rech, J. Instability of the
quantum-critical point of itinerant ferromagnets, \emph{Phys. Rev. Lett}.%
\textbf{\ 2004}, \emph{92}, 147003.

\bibitem{other2} de Oliveira, T. R.; Rigolin, G.; de Oliveira, M. C.;
Miranda, E. Multipartite entanglement signature of quantum phase
transitions. \emph{Phys. Rev. Lett}. \textbf{2007}, \emph{97}, 170401.

\bibitem{other3} Mazzanti, F.; Astrakharchik, G. E.; Boronat, J.;
Casulleras, Off-diagonal ground-state properties of a one-dimensional gas of
Fermi hard rods. J. \emph{Phys. Rev. A} \textbf{2008}, \emph{77}, 043632.

\bibitem{other4} Zhao, J.-H.; Zhou, H.-Q. Singularities in ground-state
fidelity and quantum phase transitions for the Kitaev model. \emph{Phys.
Rev. B} \textbf{2009}, \emph{80}, 014403.

\bibitem{other5} Yao, Y.; Li, H. W.; Zhang, C.-M.; Yin, Z. Q.; Chen, W. C.;
Guo, G.-C.; Han Z.-F. Performance of various correlation measures in quantum
phase transitions using the quantum renormalization-group method. \emph{%
Phys. Rev. A} \textbf{2012}, \emph{86}, 042102.

\bibitem{droplet1} Cabrera, C. R., Tanzi, L., Sanz, J., Naylor, B., Thomas,
P., Cheiney, P., Tarruell, L. Quantum liquid droplets in a mixture of
Bose-Einstein condensates.  \emph{Science} \textbf{2018}, \emph{359},
301-304.

\bibitem{droplet2} Cheiney, P., Cabrera, C. R., Sanz, J., Naylor, B., Tanzi,
L., Tarruell, L. Bright soliton to quantum droplet transition in a mixture
of Bose-Einstein condensates. \emph{Phys. Rev. Lett}. \textbf{2018}, \emph{%
120}, 135301.

\bibitem{droplet3} Semeghini, G., Ferioli, G., Masi, L., Mazzinghi, C.,
Wolswijk, L., Minardi, F., Modugno, M., Modugno, G., Inguscio, M., Fattori,
M. Self-bound quantum droplets in atomic mixtures. arXiv:1710.10890.

\bibitem{3-body1} Abdullaev, F. Kh.; Gammal, A., Tomio, L.; Frederico, T.
Stability of trapped Bose-Einstein condensates. \emph{Phys. Rev. A} \textbf{%
200}1, \emph{63}, 043604.

\bibitem{3-body2} Abdullaev, F. Kh.; Salerno, M. Gap-Townes solitons and
localized excitations in low-dimensional Bose-Einstein condensates in
optical lattices. \emph{Phys. Rev. A} \textbf{2005}, \emph{72}, 033617.

\bibitem{Shlyap} Petrov, D. S., Holzmann, M., Shlyapnikov G. V.
Bose-Einstein condensation in quasi-2D trapped gases. \emph{Phys. Rev. Lett}%
. \textbf{2000}, \emph{84}, 2551-2555.

\bibitem{Salasnich} Salasnich, L., Parola, A., Reatto, L. Effective wave
equations for the dynamics of cigar-shaped and disk-shaped Bose condensates.
\emph{Phys. Rev. A} \textbf{2002}, \emph{65}, 043614.

\bibitem{Delgado} Mu\~{n}oz Mateo, A., Delgado, V. Effective mean-field
equations for cigar-shaped and disk-shaped Bose-Einstein condensates. \emph{%
Phys. Rev. A} \textbf{2008}, \emph{77}, 013617.

\bibitem{Fermi} Giorgini, S.; Pitaevskii, L. P.; Stringari, S. Theory of
ultracold atomic Fermi gases. \emph{Rev. Mod. Phys}. \textbf{2008}, \emph{80}%
, 1215-1273.

\bibitem{Minguzzi} Capuzzi, P.; Minguzzi, A.; Tosi, M. P. Collective
excitations of a trapped boson-fermion mixture across demixing.\emph{\ Phys.
Rev. A} \textbf{2003}, \emph{67}, 053695.

\bibitem{SKA} Adhikari, S. K. Fermionic bright soliton in a boson-fermion
mixture. \emph{Phys. Rev. A} \textbf{2005}, \emph{72}, 053608.

\bibitem{Luca} Manini, N.; Salasnich, L. Bulk and collective properties of a
dilute Fermi gas in the BCS-BEC crossover. \emph{Phys. Rev. A} \textbf{2005}%
, \emph{71}, 033625.

\bibitem{Bulgac} Bulgac, A.; Local-density-functional theory for superfluid
fermionic systems: The unitary gas. \emph{Phys. Rev. A} \textbf{2007}, \emph{%
76}, 040502(R).

\bibitem{Santos} G\'{o}ral, K.; Santos, L. \emph{Phys. Rev. A} \textbf{2002}%
, \emph{66}, 023613.

\bibitem{misc0} Deconinck, B.; Kevrekidis, P. G., Nistazakis, H. E.;
Frantzeskakis, D. J. Linearly coupled Bose-Einstein condensates: From Rabi
oscillations and quasiperiodic solutions to oscillating domain walls and
spiral waves, \emph{Phys. Rev. A} \textbf{2004}, \emph{70}, 063605.

\bibitem{misc2} Adhikari, S. K.; Malomed, B. A. Two-component gap solitons
with linear interconversion. \emph{Phys. Rev. A} \textbf{2009},\emph{79},
015602.

\bibitem{misc1} Wen, L.; Liu, W. M.; Cai, Y.; Zhang, J. M.; Hum J.
Controlling phase separation of a two-component Bose-Einstein condensate by
confinement. \emph{Phys. Rev. A} \textbf{2012}, 85, 043602.

\bibitem{Mineev} Mineev, V. P. Theory of solution of two almost perfect Bose
gases, Zh. Eksp. Teor. Fiz. \textbf{1974}, \emph{67}, 263-272 [Sov. Phys.
JETP \textbf{1975}, \emph{40}, 132].

\bibitem{fluids} Wang, J.; Cerdeiri\~{n}a, C. A.; Anisimov, M. A.; Sengers,
J. V. Principle of isomorphism and complete scaling for binary-fluid
criticality. \emph{Phys. Rev. E} \textbf{2008}, \emph{77}, 031127.

\bibitem{Thawatchai} Roeksabutr, A.; Mayteevarunyoo, T.; Malomed, B. A.
Symbiotic two-component gap solitons. \emph{Opt. Exp}. \textbf{2012}, \emph{%
20}, 24559-24574.

\bibitem{GEA} Astrakharchik, G. E.; Malomed, B. A. Quantum versus mean-field
collapse in a many-body system, \emph{Phys. Rev. A} \textbf{2015}, \emph{92}%
, 043632.

\bibitem{Metropolis} Metropolis, N.; Rosenbluth, A. W.; Rosenbluth, M. N.;
Teller, A. H. Equation of state calculations by fast computing machines.
\emph{J. Chem. Phys}. \textbf{1953}, \emph{21}, 1087.

\bibitem{Jastrow} Jastrow, R. Many-body problem with strong forces. \emph{%
Phys. Rev}. \textbf{1955}, \emph{98}, 1479-1484.
\end{thebibliography}
\end{document}